**The NEOWISE-Discovered Comet Population and the CO+CO$_2$ production rates.**

James M. Bauer[1,2], Rachel Stevenson[1], Emily Kramer[1], A. K. Mainzer[1],Tommy Grav[3], Joseph R. Masiero[1], Yan R. Fernández[4], Roc M. Cutri[2], John W. Dailey[2], Frank J. Masci[2], Karen J. Meech[5,6], Russel Walker[7], C. M. Lisse[8], Paul R. Weissman[1], Carrie R. Nugent[1], Sarah Sonnett[1], Nathan Blair[2], Andrew Lucas[2], Robert S. McMillan[9], Edward L. Wright[10], and the WISE and NEOWISE Teams

[1]Jet Propulsion Laboratory, California Institute of Technology, 4800 Oak Grove Drive, MS 183-401, Pasadena, CA 91109 (email: bauer@scn.jpl.nasa.gov)

[2]Infrared Processing and Analysis Center, California Institute of Technology, Pasadena, CA 91125

[3]Planetary Science Institute, 1700 East Fort Lowell, Suite 106, Tucson, AZ 85719-2395

[4]Department of Physics, University of Central Florida, 4000 Central Florida Blvd., P.S. Building, Orlando, FL 32816-2385

[5]Institute for Astronomy, University of Hawaii, 2680 Woodlawn Dr., Manoa, HI 96822

[6]NASA Astrobiology Institute, Institute for Astronomy, University of Hawaii, Manoa, HI 96822

[7]Monterey Institute for Research in Astronomy, 200 Eighth Street, Marina, CA 93933

[8] Applied Physics Laboratory, Johns Hopkins University, 11100 Johns Hopkins Road Laurel, MD 20723---6099

[9]Lunar and Planetary Laboratory, University of Arizona, 1629 East University Blvd., Kuiper Space Science Bldg. 92, Tucson, AZ 85721-0092,

[10]Department of Physics and Astronomy, University of California, PO Box 91547, Los Angeles, CA 90095-1547

*Submitted to Astrophysical Journal May 1, 2015, revised September 2, 2015.*

**Abstract:** The 163 comets observed during the WISE/NEOWISE prime mission represent the largest infrared survey to date of comets, providing constraints on dust, nucleus sizes, and CO+CO$_2$ production. We present detailed analyses of the WISE/NEOWISE comet discoveries, and discuss observations of the active comets showing 4.6 µm band excess. We find a possible relation between dust and $CO+CO_2$ production, as well as possible differences in the sizes of long and short period comet nuclei.

## 1   Introduction

When the Wide-field Infrared Survey Explorer (WISE) mission was launched on 14 December 2009, the complete sky had not been surveyed at thermal infrared (IR) wavelengths since IRAS. The primary purpose of the WISE mission was to conduct an all-sky survey at 3.4, 4.6, 12, and 22 µm (referred to as W1, W2, W3, and W4) at unprecedented sensitivity and spatial resolution (Wright et al. 2010). An enhancement to the WISE mission was funded by NASA's Planetary Science Division, called NEOWISE, to detect moving objects in the data and to develop a searchable archive of moving object photometry and images to facilitate precovery and analysis of subsequent discoveries (Mainzer et al. 2011c, 2012a). Both aspects of NEOWISE were successful, with over 158,000 small bodies observed including 34000 discoveries. More than 616 NEOs were detected (Mainzer et al. 2012a) during the prime mission, from January 14, 2010 through February 1, 2011. NEOWISE has provided the largest catalog of thermal-infrared solar-system object data to date. The observations have yielded an unprecedented number of size measurements for a wide array of classes of solar system bodies using radiometric modeling techniques (cf. Bauer et al. 2013, Bauer et al. 2012a, Bauer et al. 2012b, Bauer et al. 2011, Mainzer et al. 2011a,  Mainzer et al. 2011b, Mainzer et al. 2011c, Masiero et al. 2011, Masiero et al. 2012, Grav et al. 2011, Grav et al. 2012). However, NEOWISE also observed the largest number of comets to date in the IR; a total of 163 comets have been identified in the data, a sample that offers a unique set of constraints on cometary physical properties. In addition to measuring the nucleus size distribution of comets, the data are used to quantify dust characteristics and mass loss, as well as gas production of rarely-observed species (Bauer et al. 2011, Bauer et al. 2012b; Stevenson et al. 2014, Stevenson et al. 2015).

The WISE/NEOWISE survey began regular survey operations on 14 January 2010 (Modified Julian Date [MJD] 55210). The secondary cryogen reservoir of solid hydrogen was depleted on 4 August 2010 (MJD 55412), resulting soon after in the saturation of the W4 channel. The survey then continued in W1-3, the so-called 3-band cryogenic phase, until the primary reservoir was depleted at the end of September 2010 (MJD 55469). After this, science survey operations were extended for the next 4 months in the W1 and W2 until 1 February 2011 (MJD 55593), when the "post-cryogenic" mission phase ended (Mainzer et al. 2012, Masiero et al. 2012). At this point the spacecraft was placed into a hibernation state. The success of NEOWISE in this first period, a little more than a year of survey operations referred to as the "prime mission", led to the decision to restart the WISE spacecraft and the survey in 2013 exclusively for the purposes of surveying solar system bodies. The reactivated spacecraft was renamed NEOWISE, after the planetary mission, and the survey has been underway since 23 December 2013 (MJD 56649; Mainzer et al. 2014). Since the reactivation, NEOWISE has detected > 12000 minor planets, including 260 NEOs and 63 comets at 3.4 and 4.6 µm.

*1.1    WISE/NEOWISE Cometary Discoveries:*

During the cryogenic mission, NEOWISE was the most prolific discoverer of comets, other than the sun-grazing comets observed by SOHO. NEOWISE discovered 18 comets during the prime mission and discovered activity on an additional three small bodies. Since the beginning of the reactivated mission, NEOWISE has discovered four additional comets.[1] For the prime mission discoveries, about half of the comets are designated long-period comets (LPCs; comets with orbital periods >200 years). For the reactivated mission half of the comets discovered are LPCs.

The new NEOWISE comets (see Figure 1) form an interesting population that has been discovered based on their thermal emission in the infrared, rather than reflected visible light. This is particularly important as the low albedos of the nuclei (Lamy et al. 2004) and potentially the darker refractory grains (Bauer et al. 2012b) make discovery in reflected light difficult until cometary activity increases the brightness dramatically upon approach to perihelion. The large-grain dust component may be comprised of dark, refractory grains that facilitate detection and study at IR wavelengths out to greater distances than can be reached by reflected light. Finally, strong gas emission lines of CO (4.67 µm) and $CO_2$ (4.23 µm) fall within the NEOWISE 4.6 µm channel (≥80% peak throughput from 4.13 to 5.14 µm; Wright et al. 2010), allowing abundance constraints to be set on these species.  CO is otherwise only observable from the ground for bright comets, or if the comet's geocentric velocity is large enough that the comet lines are sufficiently Doppler shifted from their telluric counterparts (cf. Dello Russo et al. 2009). Emitted $CO_2$ is only detectable directly from space (Bockelee-Morvan et al. 2004). In this paper, we describe this NEOWISE-discovered population in detail, including analysis of the dust, constraints on the nucleus sizes, and gas production rates of various species. We provide a wider context for the $CO+CO_2$ analyses by exploring the $CO+CO_2$ production in the full comet sample from the 163 comets, roughly a quarter of which show 4.6 μm band excess attributable to CO or $CO_2$ gas emission. Because the NEOWISE W2 band encompasses both CO and $CO_2$ features, it is difficult to separate their relative contributions; however, CO is generally more than a factor of 11 times weaker than the $CO_2$ feature (see section 4.6).

---

[1] 2010 KG43 is not included in this tally, since while activity has been reported (Waszczac et al. 2013), it has not been designated as a comet yet. The NEOWISE observations of this body are discussed in that reference. On 15 May, 2015, the NEOWISE reactivated mission discovered its 4th comet, P/2015 J3.

### 1.2  $CO+CO_2$ production rates:

Where $H_2O$-driven sublimation begins beyond 6 AU and can lift optically detectable sub-micron dust, comets are variable objects that become obviously active typically somewhere inside 4 AU when they cross the point at which water-ice sublimation becomes the dominant driver of activity (Meech & Svøren, 2004). However, the exact details of when and how active they will become remains difficult to predict as these events are sensitive to variations in their compositions. In the outer solar system, water-ice is very common, yet other common ices exist as well that can sublimate rapidly at distances greater than 4 AU. For the last several decades, comets have been grouped into dust-rich and gas-rich categories that may not necessarily correlate with their dynamical age or origin (A'Hearn et al. 1995). In most circumstances, water-ice sublimation likely drives their activity near perihelion, but at larger distances other common volatile constituents like CO and $CO_2$ may be the primary driver. Recent studies have shown that the CO or $CO_2$ production rate relative to $H_2O$ increases with heliocentric distance (A'Hearn et al. 2012), but these analyses are based on a limited sample. Some in-situ measurements, for example with comet 103P/Hartley 2 (cf. A'Hearn et al. 2011), suggest different source regions for $CO_2$ and $H_2O$ on the surface. To date, only 40 comets have had their CO or $CO_2$ production rates constrained from space-based observations (cf. Bockelee-Morvan et al. 2004, Pittichova et al. 2008, Ootsubo et al. 2012, Reach et al. 2013, Bauer et al. 2011 & 2012b). The NEOWISE sample represents a uniform survey of CO+$CO_2$ production collected with a single space-based instrument with consistent instrumental response. This sample nearly doubles the sample of measured CO+$CO_2$ production rates in comets reported in the literature. Moreover, the 12 and 22 μm channel observations set firmer constraints on the nucleus and dust contributions to the signal than do 2-band constraints such as those provided by Spitzer Space Telescope (SST; Reach et al. 2013), allowing the gas contribution to be separated.

## 2   Observations

*2.1   WISE spacecraft observations.*

During the fully cryogenic portion of the mission, simultaneous exposures in the four *WISE* wavelength bands were taken once every 11 s, with exposure durations of 8.8 s in *W3* and *W4,* and 7.7 s in *W1* and *W2* (Wright et al. 2010). The number of exposures acquired for each moving object depends on its rate of motion across the sky, as well as the rate of survey progression. A total of 8 exposures were collected for areas on the sky on the ecliptic on average at each pass, rising to several hundreds of exposures near the ecliptic poles. For most moving objects, this cadence resulted in collecting ~12 exposures uniformly spaced over ~36 hours (Mainzer et al. 2011a; Cutri et al. 2012). Note that *WISE* may have observed a subset of its full sample of observations of any particular solar system object while it was in different parts of the sky, i.e., when several weeks or months had passed since the previous exposure (e.g., comet 67P; Bauer et al. 2012b), often providing data at different viewing geometries. Henceforth, we refer to the series of exposures containing the object in the same region of sky as a "visit", or "epoch". The spatial resolution in the *WISE* images varies with the wavelength of the band. The FWHM of the mean point-spread-function (PSF), in units of arcseconds was 6.1, 6.4, 6.5, and 12.0 arcsec for *W1, W2, W3,* and *W4,* respectively (Wright et al. 2010; Cutri et al. 2012).

As with the comets we have previously studied (Bauer et al. 2011, 2012a, 2012c), some analysis was improved by stacking at the objects' rates of motion to increase the signal-to-noise ratio (SNR). For each body, the images were identified using the *WISE* image server (http://irsa.ipac.caltech.edu/applications/wise), as described by Cutri et al. (2012). Images were stacked using the moving object routine, "A *WISE*

Astronomical Image Co-adder" (AWAIC; Masci & Fowler 2009). Figure 1 shows the variation in morphology of the subset of cometary objects discovered by WISE/NEOWISE.

During the fully cryogenic prime mission, 163 comets were detected by WISE/NEOWISE with an SNR ≥ 5 in the stacked image from at least one band. Of these, 94 were detected with an SNR ≥ 5 in single-exposure images by the WISE moving-object pipeline sub-system (WMOPS). The additional 69 comets were found by co-adding the exposures at each visit. Of the comets detected, 57 were LPCs and 106 were short-period comets (SPCs; comets with orbital periods <200 years), according to present designations.[2]

We report here on the comets discovered by WISE/NEOWISE. We also discuss the total sample of comets that show 4.6 μm excess, likely attributable to $CO+CO_2$ emission. A summary of their WISE/NEOWISE observations are shown in Table 1. Note that LPCs are indicated by a "C/" prefix to their designations.

---

[2] JPL's Horizon's ephemeris service; http://ssd.jpl.nasa.gov

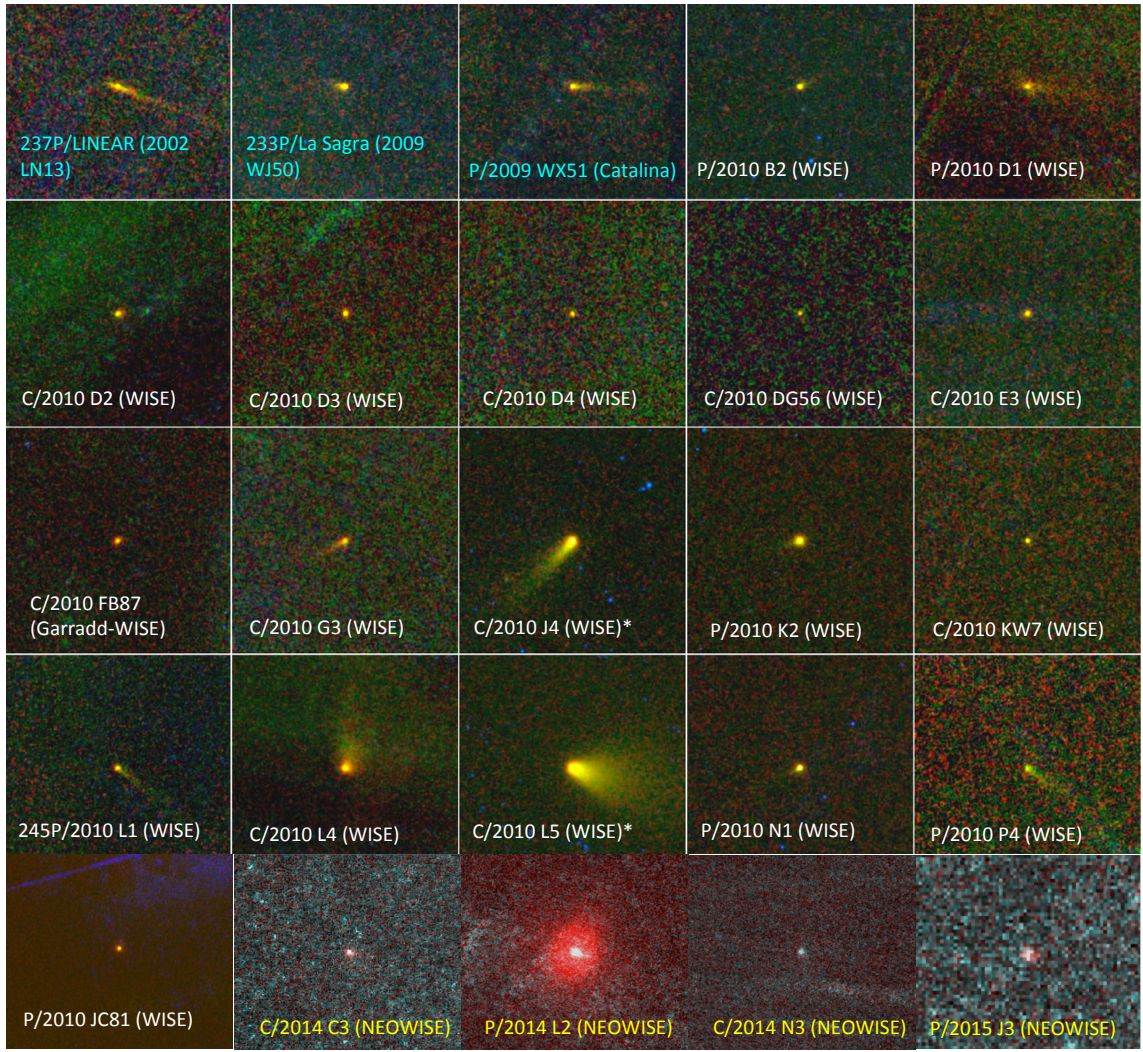

Figure 1: Discovery images of WISE/NEOWISE comets shown in 3-colors. The prime mission comets have the 22 μm image mapped to the image's red channel, the 12 μm image mapped to green, and the 4.6 μm image mapped to blue. The comets for which the activity, and not the object, were discovered by NEOWISE are shown with blue labels, in the upper left. In the case of the four comets discovered to date by the NEOWISE Reactivation (yellow text labels, on the bottom row), the 4.6 μm image is mapped to red, and the 3.4 μm image to both green and blue. The images are 6 arcmin on a side. The comets span a wide range of morphologies and activity levels; over half are LPCs.

## 2.2   Discovery Objects

As of June, 2015, there were a total of 25 cometary body discoveries made by data from the WISE spacecraft. These include three distinct categories. Comets 237P, 233P, and P/2009 WX51 (Catalina) were known objects at the time of the discovery of their activity, but were not known to be previously active. NEOWISE reported coma and trails for these objects as they were imaged during the prime mission. Additionally, WISE discovered 18 new comets during the prime mission, which were named for the spacecraft discovery.  These two groups represent a significantly different sample apart from other cometary discoveries, since each comet, or its active nature, was first discovered at thermal IR wavelengths, while comets discovered from ground-based telescopes are selected based on optical observations. Note that this sample could include a further member, 2010 KG43, a body on a centaur-like orbit that was reported to have activity when viewed by the Palomar Transient Survey (Wasczac et al. 2013). The WISE discovery observations of this object, taken at a significantly different epoch, showed no coma or extended emission.

In the first year, the reactivated NEOWISE mission has discovered three new active comets.  These comets (each called NEOWISE) were discovered from their 4.6 µm signal, and so may have yet a different set of selection biases apart from those found in the prime mission or ground-based searches.  A fourth comet, P/2015 J3 (NEOWISE) was discovered on 15 May, 2015, after the first submission of this manuscript.

## 2.3   Comets with significant 4.6 µm signal

Throughout the fully cryogenic portion of the WISE/NEOWISE mission most comets exhibited their highest signal-to-noise ratio in the 12 and 22 μm channels, as the dust, often dark and composed of refractory grains (cf. Bauer et al. 2011 and Bauer et al 2012b), provided strong thermal signal relative to the background. However, a total of 56 comets showed some signal in the 4.6 μm channel, and often also at 3.4 μm. While dust thermal emission dominates the 12 and 22 μm bands, the 3.4 μm channel is dominated by the reflected light of the dust. Weak molecular emission lines, primarily from O-H and C-H related species, fall within this channel, but this signal typically is significantly less than that of the dust signal, i.e. ~30% or less of the total signal (cf. Bockelee-Morvan 1995, Reach et al. 2013). However, strong molecular emission lines of CO (4.67 μm) and $CO_2$ (4.23 μm) exist within the 4.6 μm bandpass (cf. Pittichova et al. 2008, Bauer et al. 2011, and Reach et al. 2013). The CO and $CO_2$ emission bands are strong enough to manifest excess flux within the 4.6 μm channel, apparent when the dust signal contribution is constrained by the 3.4 μm signal and the 12 and 22 μm thermal flux. Often, there are additional morphological differences between the 4.6 μm signal and the other bands. Moreover, the shape of the comets in the 3.4 μm channel often matches better the 12 and 22 μm signal, likely attributable to dust (Figure 2).

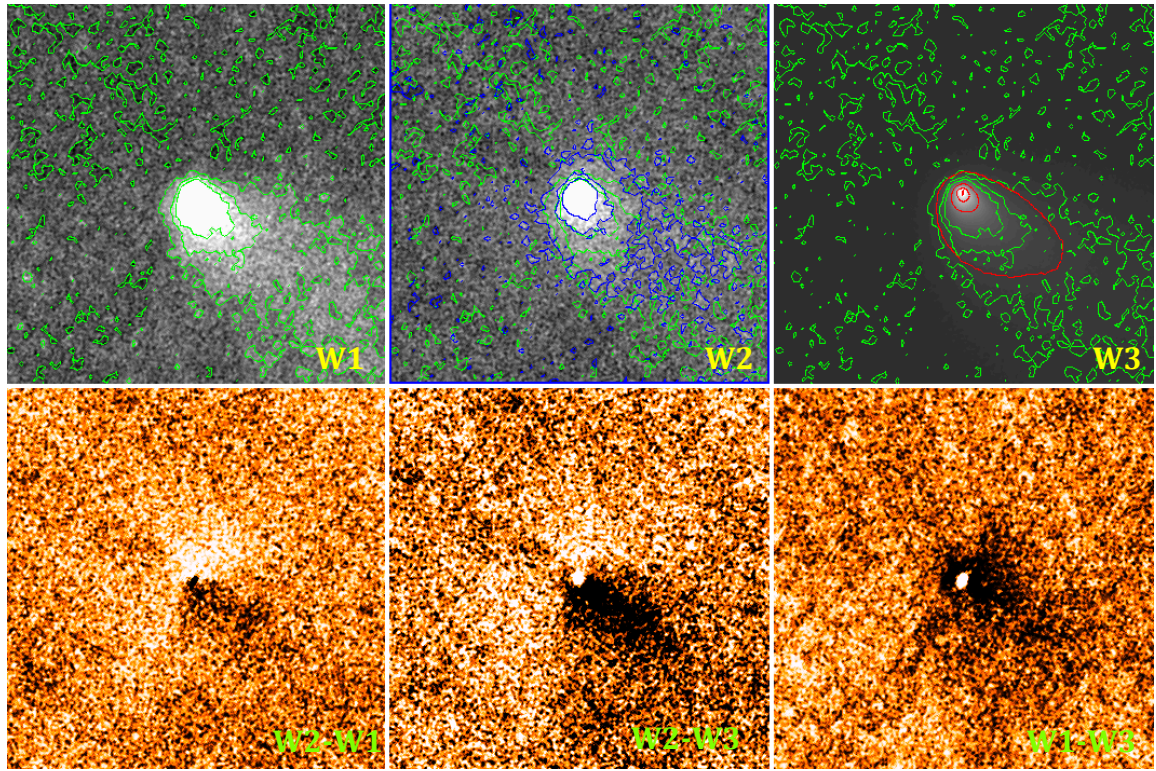

Figure 2: Morphological differences for comet C/2007 Q3 (Siding Spring); see Tables 1 and 2 for observation times and distances . The top three panels (left to right) show 3.4 μm image contours (green) overlaid onto W1, W2 (blue contours), and W3 (red contours) band images. The bottom panels show (from left to right) the peak-normalized difference images of W2-W1, W2-W3, and W1-W3. Note the miss-match between shape of the contours of W2 and W3 in the top panels, and the better match between the contours of W1 and W3. Also, note the asymmetries in the difference images for W2 that are not present in the W1-W3 image. W1 and W3 trace the dust, while a more spherical component, likely gas emission, is present in the W2 flux. Note also the point-spread function's width is larger in W3 than in W1 or W2. This is the cause of the brightness peak and more extended dark regions when W3 is subtracted from W1 in the lower right panel.

A total of 39 comets observed during the prime mission exhibited 4.6 μm band flux excess, attributable to CO to $CO_2$ emission; two of the comets, C/2009 K5 and C/2010 FB87, exhibited 4.6 μm band excess in the post-cryo mission observations. A quarter of the comets observed during the prime mission, then, exhibited 4.6 μm excess. The rate of occurrence of W2 excess differed for the comets observed thus far during the NEOWISE-Reactivation mission, which are about 2/3rds of the total (Bauer et al. 2014). However, the match is nearly identical among the prime and reactivated mission comets observed with any significant W2 signal; both samples have ~ 2/3 with 4.6 μm excess. We included in our sample of W2 excess the comets NEOWISE from the reactivated mission, three of which show 4.6 μm excess.

**Table 1: Comets Discovered by NEOWISE & with observed 4.6 μm Excess**[*]

| Comet | Ecc | q (AU) | Incl (deg) | a (AU) | $T_J$ | Earth MOID (AU) | Class | Observation MJDs (exposure mid-point times) |
|---|---|---|---|---|---|---|---|---|
| P/2010 K2 (WISE) | 0.5894 | 1.1982 | 10.642 | 2.918 | 2.972 | 0.204 | JFC | 55343.1806, 55343.3129, 55343.4452, 55343.5775, 55343.7098, 55343.8421, 55343.9744, 55344.0406, 55344.1067, 55344.1729, 55344.2390, 55344.3052, 55344.3713, 55344.4375, 55344.5036, 55344.5699, 55344.6359, 55344.7022, 55344.8345, 55344.9668, 55345.0991, 55345.2314, 55345.3637, 55345.4960 |
| P/2010 D1 (WISE) | 0.3566 | 2.6691 | 9.647 | 4.148 | 2.899 | 1.683 | JFC | 55244.3420, 55244.4742, 55244.6065, 55244.7389, 55244.7390, 55244.8713, 55244.9374, 55245.0036, 55245.1359, 55245.2021, 55245.4667, 55245.5991 |
| P/2010 D2 (WISE) | 0.4531 | 3.6590 | 57.175 | 6.691 | 1.874 | 2.944 | JFC* | 55252.6899, 55252.8223, 55252.9546, 55253.0869, 55253.2192, 55253.2854, 55253.3515, 55253.3516, 55253.5500, 55253.6823, 55253.7485, 55253.8146, 55253.8147, 55253.8808, 55253.9470, 55254.0131, 55254.2116, 55254.3439 |
| P/2010 B2 (WISE) | 0.4803 | 1.6164 | 8.931 | 3.110 | 3.013 | 0.630 | JFC (ETC) | **Epoch 1:** 55218.6376, 55218.7699, 55218.9022, 55219.0347, 55219.1670, 55219.2331, 55219.2332, 55219.4316, 55219.4978, 55219.5640, 55219.7625, 55219.8948, 55220.0271, |
| | | | | | | | | 55220.1594, 55220.1596 **Epoch 2:** 55412.2406, 55412.3728, 55412.3729, 55412.5051, 55412.5052, 55412.6374, 55412.7697, 55412.8359, 55412.9020, 55412.9682, 55413.0343, 55413.1005, 55413.1666, 55413.2328, 55413.3651, 55413.4973, 55413.4974, 55413.6296, 55413.6297, 55413.7619 |
| 245P /WISE (P/2010 L1) | 0.4663 | 2.1410 | 21.086 | 4.012 | 2.747 | 1.173 | JFC | 55349.5013, 55349.6336, 55349.7659, 55349.8982, 55349.9643, 55350.0305, 55350.0966, 55350.1628, 55350.2289, 55350.6258, 55350.7581, 55352.9409, 55352.9410, 55353.1394, 55353.2055, 55353.2717, 55353.4040, 55353.5363, 55353.6686, 55353.8009, 55353.9331, 55353.9332 |
| P/2010 N1 (WISE) | 0.5338 | 1.4945 | 12.876 | 3.206 | 2.917 | 0.491 | JFC | 55382.4788, 55382.6111, 55382.7434, 55382.8757, 55383.0080, 55383.1403, 55383.2726, 55383.3387, 55383.4710, 55383.5371, 55383.6033, 55383.6694, 55383.7356, 55383.8017, 55383.8679, 55383.9340, 55384.0001, 55384.0002, 55384.1324, 55384.3970, 55384.5293, 55384.6616 |
| 233P / La Sagra (P/2009 WJ50) | 0.4090 | 1.7950 | 11.276 | 3.037 | 3.081 | 0.818 | JFC (ETC) | 55232.9637, 55233.0960, 55233.2283, 55233.3607, 55233.6253, 55233.6915, 55233.7576, 55233.8238, 55233.9561, 55234.0223, 55234.0886, 55234.3532, 55234.4855, 55234.6179 |
| P/2009 WX51 (Catalina) | 0.7403 | 0.8000 | 9.593 | 3.080 | 2.709 | 0.009 | JFC (NEC) | 55288.5082, 55288.6405, 55288.7728, 55288.8389, 55288.8390, 55288.9051, 55288.9713, 55289.0374, 55289.1036, 55289.5005 |
| C/2010 E3 (WISE) | 1.0 | 2.2742 | 96.477 | … | … | 1.547 | LPC (Oort) | 55259.9668, 55260.0991, 55260.2314, 55260.2976, 55260.3637, 55260.4299, 55260.4960, 55260.5622, 55260.6284, 55260.6945, 55260.8268 |
| C/2010 J4 (WISE) | 1.0 | 1.0855 | 162.297 | … | … | 0.307 | LPC (Oort) | **Epoch 1:** 55317.3421, 55317.4082, 55317.4743, 55317.5404, 55317.6065, 55317.6725, 55317.8047, 55317.9370 **Epoch 2:** 55328.1127, 55328.1788, 55328.2449, 55328.3110, 55328.3771, 55328.4432, 55328.5093, |
| | | | | | | | | 55328.5754, 55328.7075 |
| C/2010 L4 (WISE) | 0.9648 | 2.8257 | 102.819 | 80.28 | -0.394 | 2.530 | LPC | 55362.4562, 55362.5885, 55362.7207, 55362.7208, 55362.7869, 55362.8530, 55362.9192, 55362.9853, 55363.0515, 55363.1176, 55363.1838, 55363.2499, 55363.3161, 55363.4484, 55363.5806, 55363.5807 |
| C/2010 L5 (WISE) | 0.9037 | 0.7908 | 147.052 | 8.208 | -0.269 | 0.114 | LPC (HTC*) | **Epoch 1:** 55361.2889, 55361.3550, 55361.5535, 55361.5536, 55361.6197 **Epoch 2:** 55391.9946, 55392.1269, 55392.2592, 55392.3254, 55392.3915, 55392.4577, 55392.5238, 55392.5900, 55392.6561, 55392.7223, 55392.8546, 55392.9869, 55393.1192, 55395.9641 |
| C/2010 D3 (WISE) | 0.9996 | 4.2476 | 76.394 | 10705 | 0.602 | 3.585 | LPC | **Epoch 1:** 55251.0085, 55251.1408, 55251.2731, 55251.5378, 55251.6701, 55252.0670, 55252.0671, 55252.1995, 55252.3318, 55252.8611, 55252.9934, 55253.1257, 55253.2580, 55253.3904, 55253.5227, 55253.6550, 55253.7874, 55254.1844, 55254.7137, 55255.1106, 55255.2430, 55255.2431, 55255.5077, 55255.6400, 55255.7723, 55255.9046, 55255.9047, 55256.0370, 55256.1693, 55256.3016, 55256.4341, 55256.6987 **Epoch 2:** 55381.5379, 55381.6702, 55381.8025, 55381.9348, 55382.2655, 55382.3317, 55382.3978, 55382.8609, 55382.9932, 55383.2578, 55383.3239, 55383.3901, 55383.5224, 55383.6547, 55383.7208, 55383.8531, 55383.9854, 55384.2500 |
| C/2010 DG56 (WISE) | 0.9764 | 1.5915 | 160.417 | 67.525 | -1.388 | 0.650 | LPC | **Epoch 1:** 55245.0010, 55245.1334, 55245.1996, 55245.2657, 55245.2658, 55245.4642 **Epoch 2:** 55403.4776, 55403.6099, 55403.6760, 55403.7422, 55403.8083, 55403.9406, 55404.0729, 55494.1250, 55494.2573, 55494.3895, 55494.3896, 55494.5218, 55494.5880, 55494.6541, 55494.7202, 55494.8525, 55494.9848 |

| Comet | | | | | | | | |
|---|---|---|---|---|---|---|---|---|
| C/2010 KW7 (WISE) | 0.9743 | 2.5704 | 147.061 | 99.85 | -1.606 | 1.626 | LPC | **Epoch 1:** 55223.9376, 55224.0699, 55224.2022, 55224.2684, 55224.3345, 55224.3346, 55224.4007, 55224.4669, 55224.5330, 55224.5992, 55224.6654, 55224.7977, 55224.9300 **Epoch2:** 55332.0641, 55332.1964, 55332.2627, 55332.3288, 55332.3950, 55332.5273, 55332.6596 |
| P/2010 JC81 (WISE) | 0.7773 | 1.8108 | 38.690 | 8.133 | 1.868 | 0.827 | HTC | 55326.6203, 55326.8849, 55327.0172, 55327.2156, 55327.2158, 55327.2819, 55327.3481, 55327.5465, 55327.6127, 55327.6788, 55327.7450, 55327.8773, 55328.0096, 55328.2742 |
| P/2010 P4 (WISE) | 0.4987 | 1.8565 | 24.102 | 3.703 | 2.740 | 0.854 | JFC | 55414.5897, 55414.7221, 55414.8544, 55414.9867, 55415.1190, 55415.1851, 55415.2511, 55415.2513, 55415.3174, 55415.3834, 55415.5157, 55415.5820, 55415.7143, 55415.8464, 55415.9787, 55416.2434 |
| 237P/LINEAR (P/2002 LN13) | 0.3526 | 2.4193 | 16.155 | 3.737 | 2.916 | 1.411 | JFC | 55357.3037, 55357.4360, 55357.5684, 55357.7007, 55357.8328, 55357.8330, 55357.8990, 55357.9651, 55358.0314, 55358.0974, 55358.1637, 55358.2297, 55358.2960, 55358.4283, 55358.5606, 55358.6929 |
| C/2010 G3 (WISE) | 0.9981 | 4.9076 | 108.268 | 2597 | -0.859 | 4.492 | LPC | **Epoch 1:** 55300.1382, 55300.2705, 55300.4028, 55300.4690, 55300.5351, 55300.6674, 55300.6676, 55300.7997, 55300.7999, 55300.8659, 55300.9320, 55300.9322, 55301.0643, 55301.0645, 55301.1306, 55301.2629, 55301.3291, 55301.3952, 55301.4614, 55301.5275, 55301.5937, 55301.6598, 55301.7260, 55301.7921, 55301.8583, 55301.9244, 55302.0567, 55302.1890 **Epoch 2:** 55380.2411, 55380.3073, 55380.3734, 55380.4396, 55380.5057, 55380.6380, 55380.7041, 55380.7703, 55380.8364 |
| C/2010 FB87 (WISE-Garradd) | 0.9905 | 2.8428 | 107.625 | 299.3 | -0.614 | 2.538 | LPC | **Epoch 1:** 55283.2198, 55283.4846, 55283.6169, 55283.6830, 55283.8154, 55283.9477, 55284.3446 **Epoch 2:** 55399.5316, 55399.5976, 55399.6639, 55399.7300, 55399.8623, |

| Comet | | | | | | | | Observation Dates (MJD) |
|---|---|---|---|---|---|---|---|---|
| | | | | | | | | 55399.9285, 55400.2592 **Epoch 3 (post-cryo):** 55573.6618, 55573.7941, 55573.9263, 55573.9264, 55573.9925, 55574.0586, 55574.1909, 55574.2570, 55574.3232, 55574.5877, 55574.8523 |
| C/2010 D4 (WISE) | 0.8894 | 7.1482 | 105.659 | 64.66 | -0.789 | 6.373 | LPC | **Epoch 1:** 55255.0178, 55255.1501, 55255.4809, 55255.5470, 55255.6795, 55255.7456, 55256.4072, 55256.5395, 55256.6719 **Epoch 2:** 55381.6953, 55381.8276, 55381.9599, 55382.0261, 55382.1583, 55382.2906, 55382.3568, 55382.4229, 55382.6214, 55382.6875, 55382.7536, 55382.8859, 55383.0182, 55383.0844, 55383.1505, 55383.2167, 55383.2828, 55383.3490, 55383.4151, 55383.4812, 55383.4813, 55383.6135, 55383.7458 |
| 2010 KG43 (undesig. periodic comet) | 0.4826 | 2.8894 | 13.616 | 5.584 | 2.695 | 1.876 | JFC (Chiron-type) | 55336.4042, 55336.4043, 55336.5365, 55336.5366, 55336.6688, 55336.6689, 55336.8011, 55336.8673, 55336.9334, 55336.9996, 55337.0657, 55337.1319, 55337.1980, 55337.2642, 55337.3965, 55337.5288, 55337.6611 |
| C/2014 C3 (NEOWISE) | 0.9828 | 1.8620 | 151.783 | 108.5 | -1.437 | 0.866 | LPC | **(re-activated)** 56702.7097, 56702.8415, 56702.9073, 56702.9731, 56703.0390, 56703.1708, 56703.3024 |
| P/2014 L2 (NEOWISE) | 0.6464 | 2.2344 | 5.1844 | 6.32 | 2.498 | 1.223 | JFC | **(re-activated)** 56815.4086, 56815.5402, 56815.6718, 56815.8034, 56815.9350, 56816.0007, 56816.0666, 56816.1324, 56816.1982, 56816.2639, 56816.3298, 56816.3956, 56816.4613, 56816.5271, 56816.6588, 56816.7904, 56816.9219, 56816.9220, 56817.0536 |
| C/2014 N3 (NEOWISE) | 0.9999 | 3.8774 | 61.642 | 40131. | 1.160 | 2.884 | LPC (Oort Cloud comet) | **Epoch 1 (re-activated):** 56841.6682, 56841.7997, 56841.9314, 56842.2603, 56842.3261, 56842.3918, 56842.4577, 56842.5235, 56842.5892, 56842.6550, 56842.7209, 56842.7866, 56842.9182, 56843.0498, 56843.1813 **Epoch 2(re-activated):** 57003.7988, 57003.9302, |

| Comet | | | | | | | | Observations |
|---|---|---|---|---|---|---|---|---|
| | | | | | | | | 57004.0618, 57004.1932, 57004.2589, 57004.3246, 57004.3903, 57004.4562, 57004.5219, 57004.6533, 57004.7847, 57004.9162, |
| P/2015 J3 (NEOWISE) | 0.5538 | 1.4941 | 8.125 | 3.348 | 2.876 | 0.499 | JFC | **(re-activated)** 57157.0565, 57157.1877, 57157.1878, 57157.3190, 57157.4502, 57157.5815, 57157.7128, 57157.7784, 57157.8440, 57157.9097, 57157.9753, 57158.0408, 57158.0410, 57158.1065, 57158.1721, 57158.3034, 57158.4347, 57158.5660, 57158.6971, 57158.8284 |
| *Additional Comets with Tentative CO+CO$_2$ Detections Based on 4.6 μm Excess* | | | | | | | | |
| 9P/Tempel 1 | 0.5116 | 1.5334 | 10.503 | 3.140 | 2.970 | 0.519 | JFC | 55296.0844, 55296.6798, 55296.2167, 55296.3490, 55296.8121, 55296.4813, 55296.6136, 55297.0768, 55296.9444, 55296.4151, 55295.9521, 55295.9520, 55296.6797, 55296.5474 |
| 10P/Tempel 2 | 0.5372 | 1.4179 | 12.029 | 3.064 | 2.964 | 0.406 | JFC | 55313.1161, 55313.2484, 55313.3807, 55313.5130, 55313.5132, 55313.6453, 55313.6455, 55313.7778, 55313.9101, 55314.0424, 55314.1747, 55314.3070, 55314.3731, 55314.4393, 55314.5054, 55314.5716, 55314.6377, 55314.6378, 55314.7039, 55314.7700, 55314.7701, 55314.8362, 55314.9023, 55314.9024, 55314.9685, 55315.0347, 55315.1008, 55315.1670, 55315.2331, 55315.2993, 55315.4316, 55315.5640, 55315.6963, 55315.8286, 55315.9609, 55316.0932, 55316.2255, 55316.3578, 55316.4901, 55322.5102, 55322.5763, 55322.5765, 55322.6425, 55322.7086, 55322.7088, 55322.8411, 55322.9734, 55323.1057, 55323.2380, 55323.3703, 55323.5026, 55323.6349, 55323.7672, 55323.8995, 55324.0318 |
| 29P/Schwassmann-Wachmann 1 | 0.0419 | 5.7580 | 9.3761 | 6.009 | 2.985 | 4.762 | JFC | 55319.1040, 55319.2363, 55319.3686, 55319.5009, 55319.5672, 55319.6332, 55319.6995, 55319.7655, 55319.8318, 55319.9641, 55320.0964, 55320.2287 |
| 30P/Reinmuth 1 | 0.5012 | 1.8832 | 8.1227 | 3.775 | 2.838 | 0.900 | JFC | 55262.6049, 55262.7372, 55262.8695, 55262.8697, |

| | | | | | | | |
|---|---|---|---|---|---|---|---|
| | | | | | | | 55263.0020, 55263.1343, 55263.2004, 55263.2666, 55263.3327, 55263.3989, 55263.4651, 55263.5312, 55263.5974, 55263.7297, 55263.8620, 55263.9943, 55264.1266, 55264.1267, 55264.2591 |
| 65P/Gunn | 0.2607 | 2.8698 | 9.2362 | 3.882 | 2.987 | 1.857 | JFC | 55310.2078, 55310.3401, 55310.4724, 55310.6047, 55310.6709, 55310.8693, 55310.9355, 55311.0016, 55311.3325, 55311.4648, 55311.5971 |
| 67P/Churyumov-Gerasimenko | 0.6410 | 1.2432 | 7.0402 | 3.4628 | 2.746 | 0.257 | JFC | see Bauer et al. 2012b |
| 74P/Smirnova-Chernykh | 0.1488 | 3.5419 | 6.6513 | 4.161 | 3.007 | 2.558 | Encke-type | **Epoch 1:** 55214.7988, 55214.9311, 55215.0634, 55215.0635, 55215.1958, 55215.2619, 55215.3281, 55215.3943, 55215.5266, 55215.6589, 55215.7913, 55215.7914  **Epoch 2:** 55384.3289, 55384.4612, 55384.5935, 55384.7258, 55384.8581, 55384.9244 |
| 77P/Longmore | 0.3579 | 2.3107 | 24.399 | 3.599 | 2.860 | 1.316 | JFC | 55320.9284, 55321.0607, 55321.1930, 55321.3253, 55321.3914, 55321.3915, 55321.4576, 55321.5238, 55321.5899, 55321.6561, 55321.7222, 55321.7884, 55323.8392, 55323.9715, 55324.2361, 55324.3022, 55324.3023, 55324.3684, 55324.8316, 55324.9639 |
| 81P/Wild 2 | 0.5380 | 1.5931 | 3.2390 | 3.449 | 2.879 | 0.599 | JFC | **Epoch 1 (four band):** 55414.0950, 55414.2273, 55414.3596  **Epoch 2 (three band):** 55414.4919, 55414.6242, 55414.6903, 55414.7565, 55414.8226, 55414.8888, 55414.9549, 55415.0211, 55415.0872, 55415.1534, 55415.2195, 55415.3518, 55415.4841, 55415.6164, 55415.7487 |
| 94P/Russell 4 | 0.3643 | 2.2315 | 6.1847 | 3.511 | 3.003 | 1.249 | Encke-type | 55335.1146, 55335.1147, 55335.5115, 55335.5116, 55335.6438, 55335.6439, 55335.7761, 55335.7762, 55335.9084, 55335.9746, 55336.0407, 55336.1069, 55336.2392, 55339.0837, 55339.2821, 55339.3483, 55339.4806, 55339.5467, 55340.0759 |
| 100P/Hartley 1 | 0.4172 | 1.9909 | 25.662 | 3.416 | 2.851 | 0.985 | JFC | 55302.4044, 55302.4045, |
| | | | | | | | | 55302.5368, 55302.6691, 55302.9337, 55302.9998, 55303.0000, 55303.0660, 55303.1323, 55303.1984, 55303.2646, 55303.3307, 55303.4630, 55303.5292, 55304.0584, 55304.0585, 55304.1908, 55304.3231, 55304.4554 |
| 103P/Hartley 2 | 0.6938 | 1.0642 | 13.604 | 3.475 | 2.641 | 0.072 | JFC (NEO) | 55326.3456, 55326.4779, 55326.6102, 55326.7425, 55326.8748, 55326.9409, 55327.0071, 55327.0732, 55327.1394, 55327.2055, 55327.2717, 55327.3378, 55327.4040, 55327.4701, 55327.6024, 55327.7347, 55327.8670, 55327.9994 |
| 107P/Wilson-Harrington | 0.6238 | 0.9938 | 2.7824 | 2.642 | 3.082 | 0.047 | Near-Earth Comet (NEC) | 55244.6083, 55244.7408, 55244.8731, 55245.0054, 55245.1377, 55245.2700, 55245.3362, 55245.4024, 55245.4685, 55245.5347, 55245.6008, 55245.6009, 55245.6670, 55245.7332, 55245.7993, 55245.8655, 55245.9317, 55245.9978, 55246.1301, 55246.2624, 55246.2626, 55246.3949, 55246.5272, 55246.6595, 55249.7692, 55249.8354, 55249.9015, 55249.9016, 55249.9677, 55250.0339, 55250.1000, 55250.1662, 55250.2325, 55250.2985, 55250.3648, 55250.4971, 55250.6294, 55250.7617, 55250.8940, 55250.8941 |
| 116P/Wild 4 | 0.3726 | 2.1851 | 3.6077 | 3.483 | 3.009 | 1.184 | Encke-type | 55302.5333, 55302.6656, 55302.6657, 55302.7980, 55302.9303, 55303.0626, 55303.1287, 55303.1949, 55303.2610, 55303.3272, 55303.3933, 55303.5256, 55303.5257 |
| 118P/Shoemaker-Levy 4 | 0.4284 | 1.9803 | 8.5131 | 3.464 | 2.960 | 1.011 | JFC | 55277.0276, 55277.5569, 55277.7553, 55277.7555, 55277.8216, 55277.8878, 55277.9539, 55278.0201, 55280.6003, 55280.7326, 55280.8649, 55280.8651, 55280.9974, 55281.1297, 55281.1958, 55281.2620, 55281.3281, 55281.3282, 55281.3943, 55281.4605, 55281.5266, 55281.5928, 55281.7251, 55281.8574, 55281.9897, 55282.1220 |
| 143P/Kowal-Mrkos | 0.4101 | 2.5382 | 4.6897 | 4.303 | 2.863 | 1.538 | JFC | 55305.5752, 55305.8399, 55305.9722, 55306.0384, 55306.1045, 55306.3030, |

| Comet | | | | | | | | Dates |
|---|---|---|---|---|---|---|---|---|
| | | | | | | | | 55306.4353, 55306.5676, 55306.6999, 55306.8322 |
| 149P/Mueller 4 | 0.3884 | 2.6509 | 29.734 | 4.334 | 2.661 | 1.728 | JFC | 55381.6882, 55381.8205, 55381.9528, 55382.0851, 55382.2174, 55382.2835, 55382.3497, 55382.4158, 55382.4820, 55382.5481, 55382.6142, 55382.6143, 55382.6804, 55382.7465, 55383.3418, 55383.3419 |
| 169P/NEAT | 0.7669 | 0.6070 | 11.305 | 2.604 | 2.888 | 0.143 | JFC (NEO) | 55322.4118, 55322.5441, 55322.5442, 55322.6764, 55322.6765, 55322.8087, 55322.8088, 55322.9411, 55323.0072, 55323.0734, 55323.1395, 55323.2057, 55323.2718, 55323.3380, 55323.4041, 55323.5364, 55323.6687, 55323.8010 |
| 203P/Korlevic | 0.3147 | 3.1823 | 2.976 | 4.6435 | 2.912 | 2.1987 | JFC | 55248.9756, 55249.1079, 55249.1742, 55251.6222, 55251.7545, 55251.8868, 55252.0191, 55252.1516, 55252.2177, 55252.2839, 55252.3500, 55252.4162, 55252.4824, 55252.6147, 55252.7470, 55252.8793 |
| 217P/LINEAR | 0.6896 | 1.2235 | 12.882 | 3.942 | 2.549 | 0.306 | JFC (NEO) | 55271.7351, 55271.8674, 55271.9997, 55271.9998, 55272.1982, 55272.2644, 55272.3305, 55272.7275, 55272.8598, 55272.9921, 55273.1244 |
| 236P/LINEAR | 0.5088 | 1.8311 | 16.334 | 3.7275 | 2.794 | 0.891 | JFC | 55367.8870, 55367.8872, 55368.0193, 55368.1516, 55368.2839, 55368.4162, 55368.5485, 55368.6809, 55368.7469, 55368.8132, 55368.8792, 55368.9455, 55369.0115, 55369.0778, 55369.1439, 55369.2099, 55369.2101, 55369.2762, 55369.3422, 55369.3424, 55369.4085, 55369.5408, 55369.6731, 55369.8054, 55369.9377, 55370.0699, 55370.0700, 55381.5794 |
| P/2009 Q4 (Boattini) | 0.5792 | 1.3208 | 10.969 | 3.139 | 2.901 | 0.361 | JFC | 55343.3181, 55343.4504, 55343.5827, 55343.7150, 55343.8473, 55343.9135, 55343.9796, 55344.0458, 55344.1119, 55344.1781, 55344.2442, 55344.3104, 55344.4427, 55344.5750, 55344.7072*, 55344.7073, 55344.8395, 55344.8396 |
| P/2010 A3 (Hill) | 0.7322 | 1.6218 | 15.028 | 6.057 | 2.278 | 0.670 | JFC | 55211.3920, 55211.5244, 55211.6567, 55211.7890, 55211.8552, 55211.9213, 55211.9214, 55211.9875, |
| | | | | | | | | 55212.0537, 55212.1198, 55212.1200, 55212.1860, 55212.2523, 55212.3846, 55212.5169, 55212.6493 |
| P/2009 T2 (La Sagra) | 0.7690 | 1.7548 | 28.106 | 7.5960 | 2.048 | 0.822 | JFC | 55217.6080, 55217.7403, 55217.8726, 55218.0049, 55218.0050, 55218.1373, 55218.2034, 55218.2696, 55218.3358, 55218.4019, 55218.4681, 55218.5342, 55218.5344, 55218.6004, 55218.6667, 55218.7327, 55218.8652, 55218.9975, 55219.1298, 55220.7839, 55220.7840, 55220.9163, 55221.0486, 55221.1809, 55221.3133, 55221.3134, 55221.3795, 55221.4457, 55221.5118, 55221.5780, 55221.6442, 55221.7103, 55221.7765, 55221.8426, 55221.9088, 55222.0411, 55222.1734, 55222.1735, 55222.3059, 55222.4382 |
| P/2009 Y2 | 0.6405 | 2.3392 | 29.93 | 6.5071 | 2.288 | 1.357 | JFC | 55210.0061, 55210.0721, 55210.0723, 55210.1384, 55210.2707, 55210.4030, 55210.4031, 55210.5354 |
| P/2010 A5 | 0.6643 | 1.7120 | 5.784 | 5.1001 | 2.493 | 0.711 | JFC | 55215.8599, 55215.9922, 55216.1245, 55216.2570, 55216.3893, 55216.4554, 55216.4555, 55216.5216, 55216.5878, 55216.6539, 55216.7201, 55216.7863, 55216.8524, 55216.9848, 55217.1172, 55217.2495, 55217.3818 |
| P/2010 H2 (Vales) | 0.1929 | 3.1077 | 14.253 | 3.850 | 2.988 | 2.130 | JFC | 55383.0064, 55383.0065, 55383.1387, 55383.2710, 55383.4033, 55383.5356, 55383.6018, 55383.6679, 55383.7340, 55383.7341, 55383.8002, 55383.8663, 55383.9986, 55384.1309, 55384.2632, 55384.3955 |
| C/2005 L3 (McNaught) | 0.9996 | 5.5936 | 139.449 | 13392. | -2.228 | 4.712 | "Comet" | 55337.1708, 55340.2798, 55340.2799, 55340.4121, 55340.5444, 55340.6106, 55340.6767, 55340.7429, 55340.8090, 55340.8752, 55340.9413, 55341.0075, 55341.1398, 55341.2721, 55341.4044 |
| C/2006 S3 (LONEOS) | 1.0030 | 5.1311 | 166.033 | -1688.7 | -2.728 | 4.131 | Hyperbolic Comet | 55338.7823, 55338.9146, 55339.0469, 55339.1792, 55339.2453, 55339.2454, 55339.3115, 55339.3776, 55339.3777, 55339.4438, 55339.5099, 55339.5100, 55339.6422, 55339.6423, 55339.7745 |

| Comet | | | | | | | | |
|---|---|---|---|---|---|---|---|---|
| C/2006 W3 (Christensen) | 0.9998 | 3.1262 | 127.075 | 17989. | -1.321 | 2.299 | "Comet" | 55305.9747, 55306.1070, 55306.2393, 55306.4377, 55306.4378, 55306.5039, 55306.5701, 55306.7024, 55306.8347 |
| C/2007 G1 (LINEAR) | 1.0015 | 2.6462 | 88.359 | -1783.0 | --- | 1.843 | Hyperbolic Comet | 55240.4971, 55240.9603, 55241.0264, 55241.0926, 55241.1588, 55241.2249, 55241.2911, 55241.3572, 55241.3573, 55241.4234, 55241.4897, 55241.6220, 55241.7543 |
| C/2007 Q3 (Siding Spring) | 1.0002 | 2.2517 | 65.650 | -9454.1 | 0.767 | 1.262 | Hyperbolic Comet | **Epoch 1:** 55206.1909, 55206.2570, 55206.3232, 55206.3894 **Epoch 2:** 55349.1498, 55349.2821, 55349.4144, 55349.5467, 55349.6790, 55349.7451, 55349.7452, 55349.8113, 55349.8774, 55349.9436, 55350.0097, 55350.0759, 55350.1420, 55350.2082, 55350.2743, 55350.3405, 55350.4066, 55350.4728, 55350.5389, 55350.6051, 55350.6712, 55350.7373, 55350.7374, 55350.8035, 55350.8696, 55350.8697, 55350.9358, 55351.0019, 55351.0681, 55351.1342, 55351.2004, 55351.2665, 55351.3327, 55351.3988, 55351.4650, 55351.5311, 55351.5973, 55351.6634, 55351.7296, 55351.8618, 55351.8619, 55351.9941, 55351.9942, 55352.1264, 55352.2587 |
| C/2007 VO53 (Spacewatch) | 0.999603 | 4.8427 | 86.995 | 12194.4 | 0.143 | 4.499 | "Comet" | 55219.3447, 55219.5433, 55219.6093, 55221.0651, 55221.1974, 55221.3297, 55221.3298, 55221.3959, 55221.5944, 55221.6606, 55221.7267, 55221.7929, 55221.8590, 55221.8592, 55221.9252, 55221.9915, 55222.0575, 55222.1238, 55222.2561, 55222.3223, 55222.3884, 55222.3885, 55222.5869, 55222.7193 |
| C/2008 FK75 (Lemmon-Siding Spring) | 1.0027 | 4.5109 | 61.175 | -1670.3 | --- | 4.056 | Hyperbolic Comet | 55273.5466, 55273.5467, 55273.6790, 55273.8113, 55273.9436, 55274.0097, 55274.0759, 55274.1420, 55274.1421, 55274.2082, 55274.2744, 55274.3405, 55274.4067, 55274.4728, 55274.5390, 55274.6051, 55275.2006 |
| C/2008 N1 (Holmes) | 0.9971 | 2.7835 | 115.521 | 973.12 | -0.885 | 2.351 | "Comet" | **Epoch 1:** 55228.2909, 55228.7541, 55228.8203, |
| | | | | | | | | 55228.8864, 55229.0187, 55229.0188, 55229.1511, 55232.9888 **Epoch 2:** 55351.5219, 55351.6541, 55351.6542, 55351.7864, 55351.9187, 55351.9849, 55352.0510, 55352.1172, 55352.1833, 55352.2495, 55352.3817, 55352.3818, 55352.5140 |
| C/2008 Q3 (Garradd) | 0.9998 | 1.7982 | 140.707 | 8926.0 | -1.286 | 0.814 | "Comet" | 55307.9318, 55310.8425, 55310.8426, 55310.9087, 55310.9749, 55311.0410, 55311.1072, 55311.2395, 55311.3718, 55311.5041 |
| C/2009 F6 | 0.9975 | 1.274 | 85.765 | 512.2 | 0.113 | 0.505 | LPC | 55256.2415, 55256.3738, 55256.5061, 55256.5724, 55256.6384, 55256.7047, 55256.7708, 55256.7709, 55256.8370, 55256.9032, 55256.9693, 55257.0355, 55257.1017, 55257.2340, 55257.3001, 55257.3663, 55257.4986, 55257.5648, 55257.6309, 55257.6971, 55257.7633, 55257.8957, 55258.0280 |
| C/2009 K5 (McNaught) | 1.0008 | 1.4224 | 103.879 | -1694.6 | --- | 0.798 | Hyperbolic Comet | **(post-cryo)** 55482.4446, 55482.5768, 55482.7091, 55482.7752, 55482.8414, 55482.9075, 55482.9736, 55482.9737, 55483.0398 |
| C/2009 P1 (Garradd) | 1.0002 | 1.5513 | 106.168 | -6285.8 | -0.432 | 1.255 | Hyperbolic Comet | 55351.6232, 55354.1369, 55354.2692, 55354.5999, 55354.6661, 55354.7322, 55354.7985, 55354.8645, 55354.9968, 55355.1291, 55355.2615 |
| C/2009 U3 (Hill) | 0.9916 | 1.4144 | 51.261 | 167.88 | 0.952 | 0.868 | "Comet" | 55212.1920, 55212.3243, 55212.4568, 55212.5229, 55212.5891, 55212.6553, 55212.7214, 55212.7876, 55212.9199, 55213.0523, 55213.1846 |
| C/2010 J1 (Boattini) | 0.9538 | 1.6957 | 134.385 | 36.723 | -0.975 | 0.735 | "Comet" | **Epoch 1:** 55287.9769, 55288.1092, 55288.1093, 55288.2416, 55288.3077, 55288.3738, 55288.3739, 55288.4400, 55288.5723, 55288.7046 **Epoch 2:** 55378.1164, 55378.2487, 55378.3810, 55378.4471, 55378.5133, 55378.5794, 55378.6456, 55378.7117, 55378.8440, 55378.9763 |

*Orbital properties and observation dates of comets discovered, or with activity discovered, by WISE/NEOWISE, and of comets with noted 4.6 μm excess detected during the prime mission. Orbital properties were recorded from JPL's Small Body Database ([http://ssd.jpl.nasa.gov/sbdb.cgi](http://ssd.jpl.nasa.gov/sbdb.cgi)) on 2015-05-15. The orbital properties include the comet's orbital eccentricity (Ecc), perihelion distance (q) in AU, orbital inclination (Inc) in degrees, orbital semi-major axis (a) in AU, Minimum Earth-Orbit intersect distance (MOID) in AU, the Jupiter Tisserand parameter, and the comet's dynamical classification. Comet names are in the IAU-standard format. If an object was observed at multiple epochs these are tabulated separately in the observation dates column, as are the phases of the mission for each epoch if any were not in the fully cryogenic mission phase.

## 3  Analysis

The *WISE* image data were processed using the scan/frame pipeline, which applied instrumental, photometric, and astrometric calibrations (Cutri et al. 2012). Image stacking and photometric analysis was conducted as in previous analyses (Bauer et al. 2011, 2012a, 2012b, and 2013; Stevenson et al. 2012 and 2015). The images were visually inspected and compared to the WISE Atlas (cf. Cutri et al. 2012) to ensure there were no inertially fixed background sources. Aperture photometry was performed on the stacked images of the 25 discovered comets and the 42 additional comets with 4.6 μm signal. Aperture radii of 9, 11 and 22 arcsec were used, the aperture sizes necessary to obtain the full signal from *W3* and *W4,* the poorest resolution *WISE* bands.

### 3.1  Flux Values

The counts were converted to fluxes using the band-appropriate magnitude zero-points and 0th magnitude flux values provided in Wright et al. (2010). An iterative fitting to a black-body curve was conducted on the two long-wavelength bands to determine the appropriate color correction as listed in the same. The extracted magnitudes for the 11 arcsec aperture were then converted to fluxes (Wright et al. 2010; Mainzer et al. 2011b) and are listed in Table 2. Proper aperture corrections are required for accurate photometry (Cutri et al. 2012), in addition to the color corrections mentioned above. With these corrections, the derived magnitudes are equivalent to the profile-derived magnitudes providing there are no artifacts, saturation, or confusion with other sources in the apertures of the objects.

**Table 2: Comet Flux Measurements**

| Comet | W1 Flux (mJy) | W2 Flux (mJy) | W3 Flux (mJy) | W4 Flux (mJy) | Apparent Activety? (Y/N/U) | Image Stack Mid-point (MJDs) |
|---|---|---|---|---|---|---|
| P/2010 K2 | 0.24+/-0.03 | 2.2+/-0.5 | 41+/-8 | 67+/-12 | Y | 55344.2734 |
| P/2010 D1 | -- | -- | 2.7+/-0.5 | 17+/-3 | Y | 55244.9688 |
| P/2010 D2 | 0.05+/-0.01 | -- | 1.9+/-0.4 | 21+/-4 | Y | 55244.9688 |
| P/2010 B2 | -- | 0.24+/-0.06 | 11+/-2 | 26+/-5 | Y | 55219.3984 |
|  | -- | -- | 1.6+/-0.3 | 4+/-1 | U | 55413.0013 |
| 245P | -- | -- | 5+/-1 | 17+/-3 | Y | 55351.7173 |
| P/2010 N1 | 0.10+/-.02 | 0.7+/-0.2 | 20+/- 4 | 42+/-8 | Y | 55383.0741 |
| 233P | 0.10+/-.02 | 0.15+/-0.04 | 9+/-2 | 24+/-5 | Y | 55233.7908 |
| P/2009 WX51 | 0.16+/-.03 | 0.7+/-0.2 | 12+/- 2 | 23+/-4 | Y | 55289.0044 |
| C/2010 E3 | -- | -- | 5 +/- 1 | 21 +/- 4 | N | 55260.4630 |
| C/2010 J4 | 0.5+/-.1 | 1.5+/-0.4 | 55 +/- 11 | 92 +/- 22 | Y | 55317.6396 |
|  | 0.4+/-.05 | 2.3+/-0.5 | 99 +/- 18 | 140+/- 25 | Y | 55328.4101 |
| C/2010 L4 | -- | -- | 7 +/- 1 | 38+/- 7 | Y | 55363.0185 |
| C/2010 L5 | 0.04 +/- .01 | 0.4+/-0.1 | 18 +/- 3 | 52 +/- 10 | Y | 55361.4543 |
|  | 0.8+/-.1 | 6+/-1 | 180+/-33 | 360+/-66 | Y | 55393.9793 |
| C/2010 D3 | 0.025+/-0.08 | -- | 0.8+/-.2 | 17+/-3 | Y | 55253.8536 55382.7616 |
|  | -- | -- | 1.8+/-.3 | 22+/-4 | Y | 55382.7616 |
| C/2010 DG56 | .13+/-.02 | .13+/-.04 | 3.0+/-.6 | 9+/-2 | N | 55245.2326 |
|  | .16+/-.02 | .8+/-.2 | 74+/13 | 145+/-27 | Y | 55403.7753 |
| C/2010 KW7 | -- | -- | 1.5+/-.3 | 7+/-2 | N | 55224.4338 |
|  | -- | -- | 5.0+/-.9 | 17+/-3 | N | 55332.3477 |
| P/2010 JC81 | 0.1+/-0.02 | 0.14+/-0.04 | 12+/-2 | 52+/-10 | N | 55327.4473 |
| P/2010 P4 | -- | -- | 4+/-.7 | 13+/-3 | Y | 55414.9204 |
| 237P | -- | -- | 4.7+/-.9 | 21+/-4 | Y | 55357.9983 |
| C/2010 G3 | -- | -- | 0.9+/-0.2 | 23+/-5 | Y | 55301.1636 |
|  | -- | -- | 2.6+/-.5 | 26+/-5 | Y | 55381.0017 |
| C/2010 FB87 | -- | -- | 2.9+/-.5 | 26+/-5 | Y | 55283.7822 |
|  | 0.9+/-.1 | 0.6+/-.1 | 12+/-2 | 142+/-26 | Y | 55399.8954 |
|  | 0.9+/-.2 | 1.2+/-.3 | -- | -- | Y | 55574.2571 |
| C/2010 D4 | -- | -- | 0.9+/-.2 | 16+/-3 | N | 55255.8449 |
| | -- | -- | 0.6+/-.1 | 18+/-4 | N | 55381.8276 |
| C/2014 C3 | 0.28+/-0.03 | 1.0+/-0.2 | -- | -- | Y | 56703.0061 |
| P/2014 L2 | 1.0+/-0.2 | 10+/-2 | -- | -- | Y | 56816.2311 |
| C/2014 N3 | 0.24+/-.06 | .37+/-.09 | -- | -- | Y | 56842.4248 |
| | 0.86+/-.19 | 1.2+/-.3 | -- | -- | Y | 57004.0486 |
| C/2015 J3 | 0.11+/-.03 | .43+/-.11 | -- | -- | U | 57157.9097 |
| **Short-Period Comets** | | | | | | |
| 9P | 0.10+/-.02 | 0.22+/- 0.06 | 18.7+/- 3.4 | 46 +/- 9 | Y | 55296.2167 |
| 10P | 3.1 +/- 0.3 | 21+/- 5 | 924+/-170 | 1109+/-203 | Y | 55314.7701 |
| 29P | 1.39+/- 0.14 | 3.17+/- 0.72 | 50+/- 9 | 1316+/-241 | Y | 55317.4994 |
| 30P | 0.91+/- 0.09 | 3.04+/- 0.69 | 277+/- 50 | 678+/-124 | Y | 55263.3989 |
| 65P | 2.10+/- 0.20 | 8.41+/- 1.89 | 209+/- 38 | 1499+/-275 | Y | 55310.9025 |
| 67P | 0.01+/-0.007 | 0.18+/-0.05 | 10.2+/-0.4 | 50+/-4 | Y | 55225.130 |
| 74P | 0.21+/- 0.03 | 0.23+/- 0.06 | 19.3+/- 3.5 | 157+/- 29 | Y | 55215.2620 |
| | 0.28+/- 0.04 | 0.39+/- 0.10 | 21.5+/- 3.9 | 196.7+/- 36.3 | Y | 55384.5936 |
| 77P | 0.10+/- 0.02 | 0.14+/- 0.04 | 15.2+/- 2.8 | 78.6+/-14.7 | Y | 55321.6562 |
| 81P | 3.63+/- 0.34 | 9.92+/- 2.22 | 446+/- 82 | 2386+/-437 | Y | 55414.2273 |
| 94P | 0.20+/- 0.03 | 0.54+/- 0.13 | 35+/- 6 | 171 +/- 32 | Y | 55337.5953 |
| 100P | 0.16+/- 0.02 | 0.28+/- 0.07 | 14 +/- 3 | 36 +/- 7 | Y | 55303.1984 |
| 103P | 0.08+/- 0.01 | 0.31+/- 0.08 | 11.1+/- 2.0 | 34.1+/- 6.5 | Y | 55327.1395 |
| 107P | -- | 0.44+/- 0.10 | 42.2+/- 7.7 | 67+/- 13 | N | 55245.9978 |
| 116P | 0.23+/- 0.03 | 0.31+/- 0.08 | 24.6+/- 4.5 | 206+/- 38 | Y | 55303.0296 |
| 118P | 1.47+/- 0.14 | 3.85+/- 0.87 | 124+/-23 | 744+/-137 | Y | 55279.5749 |
| 143P | -- | 0.09+/- 0.03 | 10.9+/- 2.0 | 57+/- 11 | N | 55306.2038 |
| 149P | 0.07+/- 0.01 | 0.12+/- 0.03 | 5.8+/-1.1 | 42+/- 8 | U | 55382.5151 |
| 169P | -- | 0.35+/-.08 | 16+/-3 | 49+/-9 | N | 55323.1064 |
| 203P | 0.27+/- 0.03 | 0.14+/- 0.04 | 20 +/-4 | 216 +/- 40 | Y | 55252.2508 |
| 217P | 0.21+/- 0.03 | 0.54+/- 0.13 | 28.4+/- 5.2 | 180+/- 33 | Y | 55272.4298 |
| 236P | 0.05+/- 0.01 | 0.21+/- 0.05 | 13.2+/- 2.4 | 25.2+/- 4.9 | Y | 55316.6008 |
| P/2009 Q4 | 0.05+/- 0.01 | 0.11+/- 0.03 | 4.0+/- 0.7 | 20+/- 4 | Y | 55344.0789 |
| P/2009 T2 | 0.41+/- 0.05 | 1.58+/- 0.36 | 130+/- 24 | 245+/- 45 | Y | 55219.1298 |
| P/2009 Y2 | 0.08+/- 0.02 | 0.16+/- 0.04 | 10+/-2 | 55.1+/- 10.4 | U | 55210.2707 |
| P/2010 A3 | 0.26+/- 0.03 | 0.77+/- 0.18 | 35.0+/- 6.4 | 143+/-26 | Y | 55212.0207 |
| P/2010 A5 | 0.26+/- 0.03 | 0.97+/- 0.22 | 100+/- 18 | 229+/-42 | Y | 55216.6539 |
| P/2010 H2 | 0.30+/- 0.04 | 0.34+/- 0.08 | 14.1+/- 2.6 | 184+/- 34 | Y | 55383.7010 |
| **Long-Period Comets** | | | | | | |
| C/2005 L3 | .16+/-.02 | .21+/-.05 | 1.2+/- 0.2 | 79+/- 15 | Y | 55340.7430 |
| C/2006 S3 | .17+/- .02 | .18+/-.05 | 1.4+/- 0.2 | 105 +/- 20 | Y | 55339.2785 |
| C/2006 W3 | 5.4 +/- 0.5 | 10.2 +/- 2.3 | 239+/- 44 | 4679+/-856 | Y | 55306.4048 |
| C/2007 G1 | 0.04+/- 0.01 | 0.11+/- 0.03 | 1.1+/- 0.2 | 48+/- 9 | Y | 55241.1257 |
| C/2007 Q3 | 14+/- 1 | 25 +/- 6 | 2670 +/- 490 | 10250+/- 1880 | Y | 55206.2567 |
| | 1.17 +/- 0.12 | 1.40 +/- 0.32 | 84 +/- 15 | 1039 +/- 190 | Y | 55350.7043 |
| C/2007 VO53 | 0.15+/- 0.02 | 0.09+/- 0.03 | 1.3+/- 0.2 | 52.+/- 10 | Y | 55221.0320 |
| C/2008 FK75 | 0.25+/- 0.03 | 0.19+/- 0.05 | 4.7+/- 0.9 | 149+/-28 | Y | 55274.3736 |
| C/2008 N1 | 0.25+/- 0.03 | 0.30+/- 0.07 | 17.2+/- 3.2 | 178+/-33 | Y | 55230.6399 |
| | 0.10+/- 0.02 | 0.17+/- 0.04 | 3.0+/- 0.6 | 52+/- 10 | Y | 55352.0180 |
| C/2008 Q3 | 0.08+/- 0.01 | 0.37+/- 0.09 | 5.9+/- 1.1 | 93+/- 17 | Y | 55309.7180 |
| C/2009 F6 | -- | 0.02+/-0.01 | 1.0+/- 0.2 | 25+/- 5 | Y | 55256.9693 |

| C/2009 K5 | 1.47+/- 0.33 | 4.46+/- 1.14 | -- | -- | Y | 55482.7423 |
|---|---|---|---|---|---|---|
| C/2009 P1 | 0.35+/- 0.04 | 0.67+/- 0.16 | 4.4+/- 0.8 | 221+/- 41 | Y | 55353.4424 |
| C/2009 T1 | 0.07+/- 0.01 | 0.06+/- 0.02 | 0.37+/- 0.07 | 27.2+/- 5.2 | Y | 55211.2139 |
| C/2009 U3 | 0.69+/- 0.07 | 2.82+/- 0.64 | 56+/- 10 | 192+/- 36 | Y | 55212.6884 |
| C/2010 J1 | 1.13+/- 0.12 | 2.97+/- 0.67 | 230+/- 42 | 521+/- 96 | Y | 55288.3078 |
|  | 0.06+/- 0.01 | 0.45+/- 0.11 | 36+/- 7 | 85+/- 16 | Y | 55378.5134 |

*Fluxes from stacked images of comets observed by WISE/NEOWISE. If an object was observed at multiple epochs these are tabulated separately in the observation dates column. Apertures of 11 arcsec in radius were used for the flux values, and the uncertainties were derived from the background noise statistics measured in the stacked images. Whether the comet had apparent coma (Y=Yes, N=No, U=Uncertain), and the mid-point times of each combined image set from each visit are listed in the table's last two columns.

## 3.2 Nucleus Sizes

In order to extract the nucleus signal for the WISE/NEOWISE comet discoveries, we used routines developed by our team (Lisse et al. 1999 and Fernandez et al. 2000) to fit the coma as a function of angular distance from the central brightness peak along separate azimuths, as applied in Bauer et al. 2011, and 2012b. As per the description in Lisse et al. (1999), the model dust coma was created using the functional form $f(\Theta) \times \rho^{-n}$, where $\rho$ is the projected distance on the sky from the nucleus and $\Theta$ is the azimuthal angle. In order to compensate for the *WISE* instrumental effects, the model coma was then convolved with the instrumental PSF appropriate for AWAIC co-added images for the matching phase of the mission (see Cutri et al. 2012). Radial cuts through an image of the comet were made every 3° in azimuth, and the best-fit radial index, *n,* and scale, *f,* at each specific azimuth were found by a least-squares minimization fit of the model to the data along that azimuth. The pixels between 5 and 20 arcsec of the brightness peak were used to fit the model coma. For most comets, the coma model fit residuals yielded uncertainty in the photometry of the extracted point-source ~10% for the W3 and W4 images; these uncertainties were similar to the photometric uncertainties in the combined nucleus and coma signals. However, higher residuals ~30% were seen for comets C/2010 L5 and C/2010 J4, and are therefore noted as possible upper limits to the nucleus sizes. We also note below that WISE imaged the predicted position of C/2010 L5 in January, before its discovery and did not detect the comet. The detection threshold is less than, but on the order of, the listed nucleus size in Table 3 (Kramer et al. 2015).

The extracted nucleus signals in W3 and W4 were fit to a NEATM model (Harris 1998, Delbo et al. 2003, and Mainzer et al. 2011b) with fixed beaming ($\eta$) parameters. The fits to only 2 extracted thermal flux points, with increased uncertainties from the raw extractions, were too poorly constrained to leave $\eta$ as a free parameter to the fit such that it converged to physically realistic values between 0.5 and 3.0. For each comet, fits were used with beaming parameter values fixed to 0.8, 1.2 (Stansberry et al. 2008), and 1.0 (Fernandez et al. 2013). Note that each attempt of a fit requires an interpolation for surface temperature in the *WISE* bands (Wright et al. 2010), so that different flux values are derived for each final fit. Table 3 presents the fit results and uncertainties, while it should be noted that, owing to the uncertainty inherent in the thermal models, there is an additional ~10% uncertainty in the derived diameter values (Mainzer et al. 2011b, 2011c). The interpolated corrections for temperature are largest in W3. Note that for P/2015 J3 (NEOWISE), the size was based on the W2 signal assuming no coma contamination for an object with a beaming parameter of 1.

Unlike the asteroids, in most cases the visual band magnitudes of the nuclei were unmeasured. Nuclei were either obscured by activity at shorter wavelengths, or were not measured at distances where they were inactive. Since only the 12 and 22 µm channel images were used (except for P/2015 J3), the albedo was relatively unconstrained. For the thermal fits of the diameters, the geometric visible albedo ($p_v$) was free, but always converged near the initial condition of a few percent. We used the beaming parameter that provided the smallest fit residuals; those $\eta$ values are listed in Table 3. An assumed 0.2 uncertainty in $\eta$ is included in the listed diameter uncertainty. However, the uncertainty in $p_v$ would have negligible effect, and so no uncertainty from that term is included with the diameter values listed in Table 3.

**Table 3: Nucleus Sizes of the Cryogenic Mission Cometary Discoveries.**

| Comet | Diameter [km] | $\eta$ | $p_v$ | Comments |
| --- | --- | --- | --- | --- |
| P/2010 D1 | 2.53+/-0.89 | 1.2 | 0.04 | |
| P/2010 D2 | 4.65+/-1.05 | 1.2 | 0.04 | |
| P/2010 B2 | 0.99+/-0.22 | 1.2 | 0.04 | |
| 245P | 1.50+/-0.33 | 1.2 | 0.04 | |
| P/2010 N1 | 0.86+/-0.26 | 0.8 | 0.04 | |
| 233P | 1.08+/-0.22 | 1.2 | 0.04 | |
| P/2009 WX51 | 0.43+/-0.10 | 1.2 | 0.04 | |
| C/2010 E3 | 1.73+/-0.36 | 1.2 | 0.04 | No coma seen during WISE observations; JPL Horizon's nucleus magnitude yields pv=0.023+/-0.01 |
| C/2010 J4 | 0.56+/-0.2 | 1.2 | 0.03 | possible upper limit |
| C/2010 L4 | 3.4+/-0.72 | 1.2 | 0.04 | |
| C/2010 L5 | 2.2+/-0.86 | 1.0 | 0.05 | possible upper limit |
| C/2010 D3 | 4.3+/-0.96 | 1.0 | 0.04 | |
| C/2010 DG56 | 1.51+/-0.27 | 1.2 | 0.04 | No coma seen; JPL Horizon's nucleus visible magnitude yields $p_v$=0.021+/-0.005 |
| C/2010 KW7 | 5.69+/-1.6 | 1.0 | 0.04 | JPL Horizon's nucleus magnitude yields pv=0.025+/-0.005 |
| P/2010 JC81 | 15.7+/-4.74 | 0.8 | 0.03 | No coma seen; JPL Horizon's nucleus visible magnitude yields $p_v$=0.03+/-0.02 |
| P/2010 P4 | 0.94+/-0.16 | 1.2 | 0.05 | |
| 237P | 2.06+/-0.34 | 1.0 | 0.04 | |
| C/2010 G3 | 7.84+/-1.44 | 1.2 | 0.04 | |
| C/2010 FB87 | 4.88+/-1.12 | 1.2 | 0.04 | |
| C/2010 D4 | 25.6+/-6.8 | 0.8 | 0.05 | No coma seen; JPL Horizon's nucleus visible magnitude yields $p_v$=0.05+/-0.02 |
| P/2010 K2 | 0.74+/-0.11 | 1.0 | 0.05 | |
| P/2015 J3 | 2.3+/-0.82 | 1.0 | 0.02 | No coma seen during NEOWISE observations; JPL Horizon's nucleus magnitude yields pv=0.02+/-0.02 |

### 3.3    Dust Photometry, Temperature, and $CO_2$/CO Production Measurements

From the total NEOWISE/WISE-observed sample of comets during the prime mission, 62 comets (or 38%) showed significant signal in W2, as shown in Tables 1 and 2. Not all comets that show significant signal in W2 have significant 4.6 μm flux excess. Some of the signal in W2 is due to the thermal and reflected light emission of the dust. In order to constrain the dust (and potentially nucleus) thermal and reflected light flux component, it is necessary to extrapolate from the thermal and shorter wavelengths. We fit a Planck function to the W3 and W4 flux, and a solar spectrum to the W1 flux for each comet. Using these two components, we determined the predicted flux at W2 (see Figure 3). We consider the flux in excess of the dust (and nucleus) signal when the flux is significantly (at the 3-sigma level) above the estimated dust and nucleus signal. We use this criterion to be certain of the excess. For those few cases (107P, 169P, and C/2009 F6) where there is weak or no significant W1 signal, we calculate the reflected light signal from the dust thermal signal and assume a corresponding emissivity of 0.9, and a surface reflectance of ~0.1 to convert this to a reflected light component. For the objects lacking thermal flux measurements (C/2014 C3, P/2014 L2, C/2014 N3, and C/2009 K5), we make the reverse assumptions and extrapolate the thermal component (Stevenson et al. 2015). Using these methods, 42 comets show W2 excess. The excess flux is then converted into $CO_2$ production estimates using the method described in Bauer et al. (2011, 2012a,b).

We noted in section 2.3 that the 3.4 µm coma signal is likely dominated by dust. However we note here that though that is probably true in most cases when the comet has a moderate to high dust-to-gas ratio, for rare gassy comets it may be that a small but significant component of the signal may be due to weak emission lines, leading to an overall significant contribution to the signal by emissions from organics and OH. Because this is the only reflected light baseline point, the 3.4 µm–based dust reflected light scaling could lead to uncertainties in the baseline fit in some cases. The result would be an underestimate of the CO+CO2 production. We also note, however, the case of C/2006 W3 (Christiansen), which had a reportedly high gas content (cf. de Val-Borro et al. 2014), yet for which we had no significant W1 signal, and derived a high CO+CO$_2$ production.

The CO$_2$ production rates ($Q_{CO2}$) are provided as a proxy for the combine CO+CO$_2$ production (see Section 4.6), and as a convenience for analysis of the behavior of the comets, with noted limitations regarding the true fraction of the CO and CO$_2$ species present. The listed uncertainties in the derived CO$_2$ production rates are the combination of two components. The first component is the uncertainty in the calculated dust contribution as constrained by the W1, W3, and W4 photometry, a comparatively small component owing to the requirement that the flux is significantly (at the 3-sigma level) above the estimated dust and nucleus signal. The second component in the listed uncertainty is from the uncertainty in the W2 signal photometry, and is added in quadrature with the dust model uncertainty. Possible systematic sources of uncertainties, such as large variations in the fraction of CO relative to CO$_2$ or contributions to the W1 flux from non-dust signal, are not included in the tabulated uncertainty values.

Lisse et al. 1998 demonstrated how broadband photometry can be applied to determine the quantity and temperature of the coma dust. As in Bauer et al. (2011, 2012a,b), we performed blackbody temperature fits to the dust coma region surrounding the nucleus, extracting the W3 and W4 measured nucleus flux contribution from the thermal signal. We calculate the effective area for the dust using 9, 11 and 22 arcsec radius apertures. For the thermal bands, this derived area has a factor of the emissivity, ε, incorporated into the result. Division by the projected length scale of the apertures, i.e., the ρ value, and by the constant π, provides an aperture-independent means of comparison to the quantity of dust visible at particular wavelengths analogous to Afρ (A'Hearn et al. 1984). We call this factor εfρ, as introduced by Lisse et al. (2002) and used in Kelley et al. (2012), which is listed in Table 4. The value of εfρ assumes that the observed flux is attributable primarily to the dust continuum emission and is the product of the emissivity, ε, the fractional area within an aperture filled by the dust, $f$, and the projected length scale of the aperture radius on the sky at the distance of the comet, ρ, expressed in centimeters. We compute our εfρ values multiplying the observed surface brightness of the comet, $I_\lambda$, by ρ and dividing by the Planck function, $B\lambda$ (Tb), where $Tb$ is the *fitted* blackbody temperature of the dust. The effective area is derived from this quantity by multiplying by πρ/ε using an assumed value of ε ≈ 0.9. Our uncertainties come from the standard deviation between the derived values for 9, 11, and 22 arcsecond apertures. Note that we do not correct εfρ values for phase angle (the Sun-target-observer angle) effects, and where we are assuming an idealized 1/$ρ$ behavior in our estimates of uncertainty, we do not find strong deviations in most cases.

The derived dust and production rates are listed in Table 4. It should be noted that for the comets without W3 or W4 signal, it is not possible to derive εfρ values directly from measurements. Therefore, we estimated these values based on the means of the (ε*fρ* - A*fρ*) values (denoted $\Delta_{(\varepsilon f\rho - Af\rho)}$) from our comet sample that had W1 and W3 or W4 signal. We found a mean value of $\Delta_{(\varepsilon f\rho - Af\rho)}$ = 0.74 +/- 0.29 from these comets; therefore, listed values for C/2009 K5, C/2010 FB87, C/2014 C3, P/2014 L2, and C/2014 N3 are simply the W1 derived $Af\rho$ values +0.74, with 0.29 added to the $Af\rho$ uncertainty in quadrature.

**Table 4: Dust and W2 excess Analysis Results**

| Visit | $R_h$ [AU] | Delta [AU] | $Q_{CO2}{}^*$[$\log_{10}$ molecules s$^{-1}$] | $\varepsilon f\rho$ [$\log_{10}$(cm)] | $Af\rho$ [$\log_{10}$(cm)] | $T_{eff}$ [K] |
|---|---|---|---|---|---|---|
| P/2009 WX51 | 1.26 | 0.76 | 25.20+/-0.12 | 0.72+/-0.07 | 0.58+/-0.08 | 241+/-7 |
| 233P | 1.80 | 1.43 | 25.04 +/-0.11 | 1.18 +/-0.10 | 0.58+/-0.09 | 213+/-1 |
| P/2010 K2 | 1.29 | 0.73 | 25.74+/-0.13 | 1.10+/-0.08 | 0.68+/-0.11 | 257+/-2 |
| P/2010 N1 | 1.55 | 1.08 | 25.49+/-0.14 | 1.19+/-0.10 | 0.67+/-0.08 | 233+/-1 |
| C/2010 L5 Epoch1 | 1.21 | 0.65 | 26.43+/-0.08 | 2.12+/-0.08 | 1.63+/-0.10 | 240+/-1 |
| C/2010 L5 Epoch2 | 1.60 | 1.15 | 25.08+/-0.08 | 1.18+/-0.08 | 0.04+/-0.06 | 213+/-3 |
| C/2010 FB87 Epoch3 | 2.92 | 2.75 | 26.48+/-0.11 | *3.4+/-0.3* | 2.64+/-0.14 | 257+/-2 |
| C/2014 C3 | 1.90 | 1.60 | 25.82+/-0.08 | *2.2+/-0.2* | 1.45+/-0.05 | 207 |
| P/2014 L2 | 2.26 | 1.98 | 27.38+/-0.08 | *3.1+/-0.3* | 2.35+/-0.10 | 195 |
| C/2014 N3 Epoch1 | 4.43 | 4.27 | 26.43+/-0.10 | *3.2+/-0.2* | 2.51+/-0.11 | 136 |
| C/2014 N3 Epoch2 | 3.96 | 3.79 | 26.51+/-0.11 | *3.6+/-0.2* | 2.78+/-0.10 | 144 |
| SPCs With 4.6 μm excess | | | | | | |
| 29P | 6.21 | 6.04 | 27.84+/-0.09 | 4.61+/-0.25 | 3.79+/-0.05 | 130+/-13 |
| 30P | 1.92 | 1.56 | 26.18+/-0.17 | 2.64+/-0.08 | 1.98+/-0.05 | 225+/-3 |
| 65P | 2.46 | 2.25 | 27.28+/-0.09 | 3.60+/-0.08 | 2.70+/-0.06 | 164+/-1 |
| 67P+ | 3.32 | 3.31 | 25.87+/-0.14 | 2.16+/-0.07 | 0.94 +/-0.26 | 183+/-4 |
| 74P Epoch 1 | 3.61 | 3.44 | 25.88+/-0.10 | 2.87+/-0.09 | 2.23+/-0.08 | 158+/-1 |
| 74P Epoch 2 | 3.74 | 3.51 | 26.22+/-0.15 | 3.04+/-0.09 | 2.36+/-0.11 | 153+/-1 |
| 77P | 2.99 | 2.80 | 25.45+/-0.15 | 2.29+/0.08 | 1.68+/-0.08 | 178+/-2 |
| 81P | 2.22 | 1.87 | 27.18+/-0.14 | 3.59+/-0.08 | 2.77+/-0.05 | 178+/-3 |
| 94P | 2.27 | 1.94 | 25.91+/-0.15 | 2.43+/-0.08 | 1.51+/-0.13 | 180+/-1 |
| 100P | 2.23 | 1.98 | 25.44+/-0.12 | 1.49+/-0.09 | 1.46+/-0.05 | 217+/-1 |
| 103P | 2.29 | 2.04 | 25.72+/-0.10 | 1.56+/-0.09 | 1.17+/-0.05 | 206+/-1 |
| 116P | 2.82 | 2.64 | 25.84+/-0.12 | 2.88+/-0.08 | 1.94+/-0.06 | 157+/-1 |
| 118P | 2.09 | 1.75 | 26.71+/-0.09 | 3.13+/-0.07 | 2.30+/-0.04 | 173+/-3 |
| 143P | 3.25 | 3.08 | 25.45+/-0.32 | 2.19+/-0.10 | 0.48+/-0.05 | 173+/-2 |
| 149P | 2.80 | 2.52 | 25.52+/-0.10 | 2.10+/-0.09 | 1.34+/-0.07 | 162+/-2 |
| 169P | 2.27 | 1.94 | 25.75+/-0.15 | 1.69+/-0.10 | 0.68+/-0.07 | 203+/-3 |
| 217P | 2.46 | 2.15 | 26.02+/-0.10 | 2.63+/-0.07 | 1.86+/-0.27 | 169+/-1 |
| P/2009 Q4 | 2.32 | 2.02 | 25.23+/-0.17 | 1.53+/-0.05 | 0.99+/-0.08 | 182+/-4 |
| P/2010 A3 | 1.85 | 1.50 | 25.86+/-0.11 | 2.15+/-0.09 | 1.37+/-0.07 | 191+/-1 |
| P/2010 H2 | 3.15 | 2.89 | 25.96+/-0.10 | 3.06+/-0.08 | 2.20+/-0.06 | 143+/-3 |

| LPCs with 4.6 µm excess | | | | | | |
|---|---|---|---|---|---|---|
| C/2005 L3 | 8.21 | 8.08 | 26.91+/-0.10 | 3.92+/-0.08 | 3.15+/-0.09 | 102+/-2 |
| C/2006 S3 | 7.23 | 7.16 | 26.55+/-0.14 | 3.93+/-0.07 | 3.05+/-0.05 | 104+/-1 |
| C/2006 W3 | 4.15 | 4.02 | 27.88+/-0.09 | 4.76+/-0.07 | 3.83+/-0.01 | 133+/-3 |
| C/2007 Q3 Epoch 1 | 2.50 | 2.25 | 27.55+/-0.10 | 4.18+/-0.08 | 3.53+/-0.05 | 192+/-2 |
| C/2007 Q3 Epoch 2 | 3.45 | 3.27 | 26.68+/-0.10 | 3.85+/-0.07 | 2.91+/-0.04 | 144+/-1 |
| C/2008 FK75 | 4.77 | 4.66 | 25.97+/-0.12 | 3.53+/-0.09 | 2.65+/-0.08 | 120+/-1 |
| C/2008 N1 Epoch 1 | 3.08 | 2.93 | 25.93+/-0.10 | 2.96+/-0.08 | 2.07+/-0.08 | 150+/-2 |
| C/2008 N1 Epoch 2 | 3.76 | 3.53 | 25.99+/-0.32 | 2.68+/-0.09 | 2.01+/-0.09 | 136+/-2 |
| C/2008 Q3 | 3.97 | 3.79 | 26.47+/-0.10 | 2.95+/-0.08 | 1.98+/-0.06 | 139+/-3 |
| C/2009 K5 | 2.65 | 2.45 | 27.06+/-0.10 | *3.35+/-0.2* | 2.63+/-0.09 | 176+/-1 |
| C/2009 P1 | 6.33 | 6.23 | 27.16+/-0.10 | 4.03+/-0.09 | 3.17+/-0.10 | 111+/-1 |
| C/2009 U3 | 1.67 | 1.29 | 26.34+/-0.10 | 2.15+/-0.08 | 1.64+/-0.05 | 201+/-2 |
| C/2010 J1 Epoch 1 | 1.84 | 1.56 | 25.94+/-0.37 | 2.48+/-0.08 | 2.07+/-0.07 | 229+/-5 |
| C/2010 J1 Epoch 2 | 2.49 | 2.20 | 25.71+/-0.16 | 1.88+/-0.09 | 1.23+/-0.08 | 222+/-3 |

[*] $Q_{CO2}$ production rates are a proxy for the combined rates derived from CO+CO2 emission (Section 3.3). [+] from Bauer et al. 2012b.

Fitting of dust and $CO_2$ excess (Figure 3).

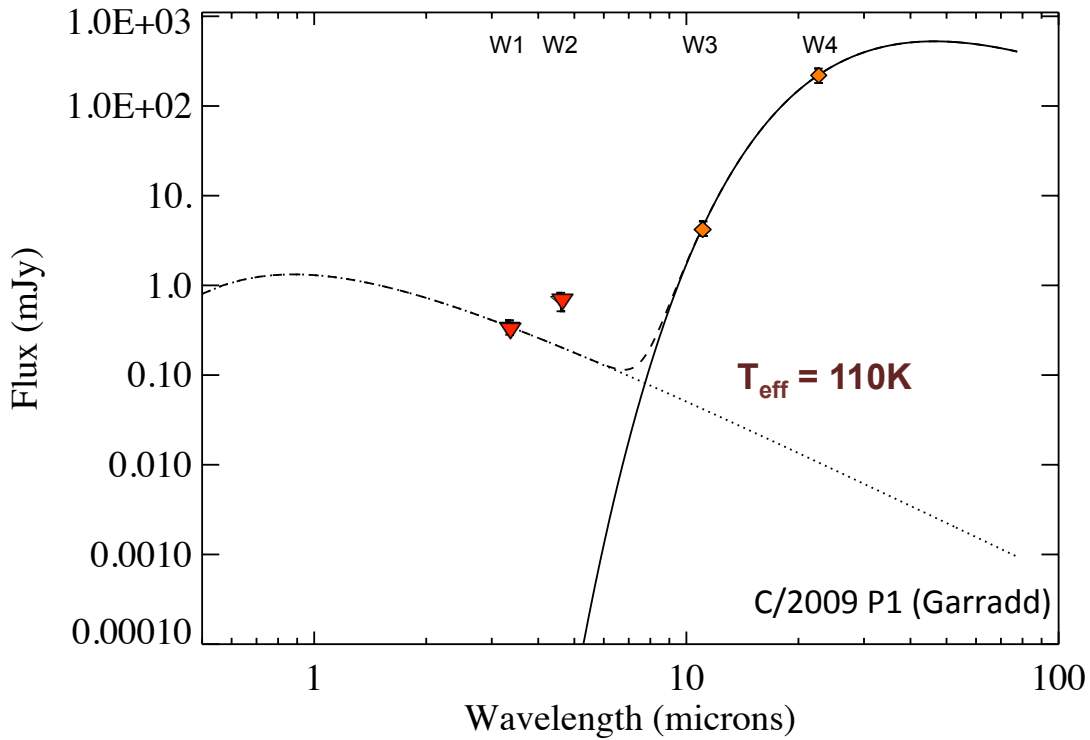

Figure 3: WISE 4.6 μm band (W2) contains CO 4.7 μm and CO2 4.3 μm emission lines. C/2009 P1 Garradd's 4.6 μm band excess not consistent with reflected or thermal contributions of coma or nucleus, but are with $CO_2$ & CO emission. The flux from the 3.4 μm (left red triangle), 4.6 μm (right red triangle), 12 μm (left orange diamond) and 22 μm (right orange diamond) channels are shown. Also the reflected light model (dotted line) thermal model (solid line) and combined signal (dashed line) are over-plotted.

## 4 Discussion

The comets observed by the WISE/NEOWISE and Reactivated NEOWISE missions represent the largest sample of comets observed in the near infrared, as summarized in Table 5. The data sets, though obtained with the same spacecraft, vary in what they offer in terms of potential measurements, owing to both the different nature of the various missions (e.g. which bands were and remain operational), and the nature of the activity in the comets themselves. For these analyses, we have focused on the initial study of the comets discovered by WISE/NEOWISE (the 21 confirmed comets), and the 3 objects whose activity were discovered by the reactivated NEOWISE mission in its first year, as well as those active comets that exhibited W2 excess during the prime mission, 39 in total, including the 9 from the cometary discoveries. Fluxes have been reported in §3.1 for all the cometary discoveries (including 2010 KG43, reported in Wasczac et al. 2013, but not yet officially designated as a comet), and the known comets with significant W2 signal taken during the WISE/NEOWISE prime mission, for a total of 56 comets, more than a third of the total WISE/NEOWISE prime mission sample.

**Table 5: Summary of Comets Observed by WISE/NEOWISE**

| Category | Total | WISE/NEOWISE Prime Mission | Reactivated NEOWISE Mission |
|---|---|---|---|
| Detected Comets | 226 | 163 | 66 |
| Detected LPCs | 86 | 57 | 29 |
| Detected SPCs | 143 | 106 | 37 |
| **Discovered Comets*** | **22** | **18+1*** | **4** |
| **Known Asteroids with Cometary Activity Discovered by NEOWISE** | 3 | 3 | -- |
| **Comets with Significant W2 signal** | 118 | **52** | 66 |
| **Comets with 4.6 $\mu$m Excess** | **39** | **36** | **3** |

NEOWISE reactivated mission tally was as of May 15, 2015. *2010 KG43, reported by Waszczak et al. 2013, no cometary designation yet, and so is excluded from the mission total.

*4.1 Thermal Dust:*

The values for $\Delta_{(\varepsilon f_\rho - Af_\rho)}$ were for active comets. The distributions of $\Delta_{(\varepsilon f_\rho - Af_\rho)}$ were similar for long and short period comets. The mean offset in the log values corresponds roughly to the a factor of 5.5 difference, which is less than, but within a factor if 1.6 of, what may be expected for a 3.4 µm albedo of ~0.1 and emissivity of ~0.9 for the same dust particles. The mean for $\Delta_{(\varepsilon f_\rho - Af_\rho)}$, however, is notably different when considering $R_h$ (see Figure 4). The number of SPCs in this sample is too small at large $R_h$ to be statistically significant. Yet, for the LPCs in our sample, for $R_h < 3$ AU, $<\Delta_{(\varepsilon f_\rho - Af_\rho)}> = 0.57 +/- 0.14$, and for $R_h < 3$ AU, $<\Delta_{(\varepsilon f_\rho - Af_\rho)}> = 0.86 +/- 0.10$. This could be indicative of larger grains being lifted by activity at greater distances rather than shorter, or possibly, and perhaps more likely, the persistence of larger grains that remain in the dust coma after peak activity.

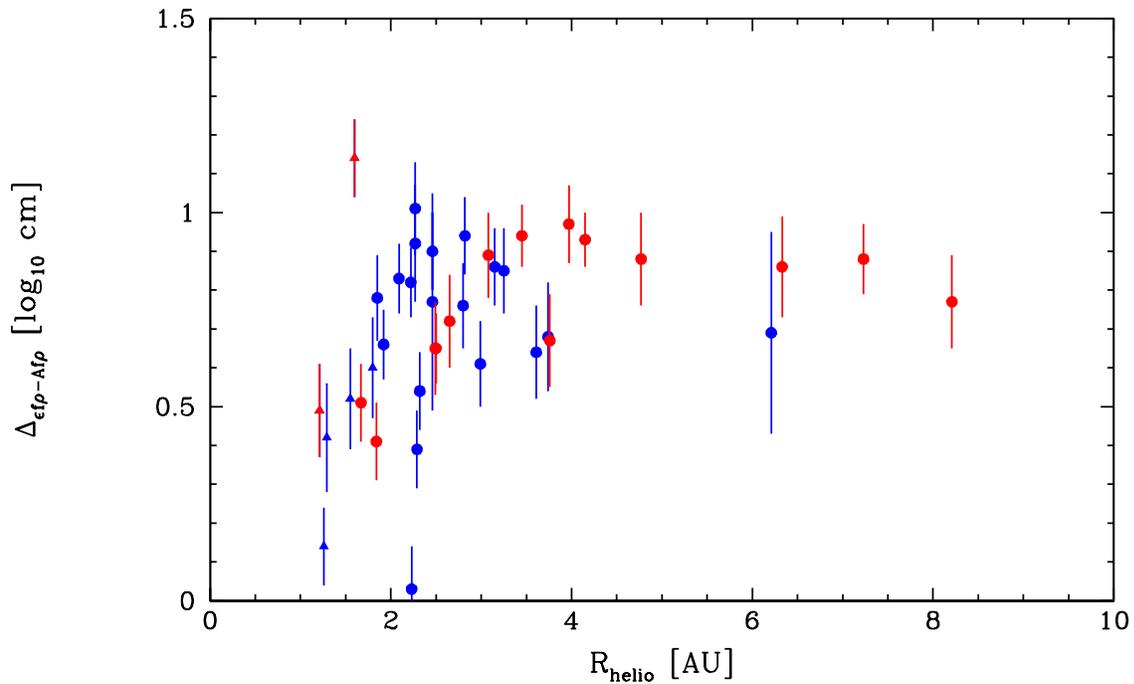

Figure 4: The difference between the log $\varepsilon f\rho$ values derived from 11 and 22 μm dust emission and the log A$f\rho$ values derived from 3.4 μm dust reflectance. LPCs (red symbols), and SPCs (blue) are show, with the WISE/NEOWISE discovered comets represented using triangles, and the remaining sample by filled circles.

*4.2 NEOWISE Discovered Comets:*

A total of 21 comets were discovered by WISE/NEOWISE during the prime mission, and 4 additional comets have been discovered during the first year of the NEOWISE reactivated mission. Of these 25 objects, 12 are designated LPCs, and those 10 observed during the prime mission have yielded constraints on their nucleus size and dust, along with $CO_2$ production rates. This gives us a good statistical basis to search for differences between SPCs and LPCs that may be attributed to formation conditions. This small set, a subset of the larger set of 163 comets observed, allows for an unprecedented comparison of nucleus sizes between SPCs and LPCs, for example, using the same methods for each comet. The same methodology does not mean there is no variation on the efficacy of the methods, however, as each comet's behavior varies greatly. Constraints on dust and $CO_2$ production require the comet be active, or recently active in the case of the dust, where the presence of strong activity during the WISE observations obviously hampers the derivation of nucleus sizes. We provide a description for each comet's behavior below (see also Figure 1).

*237P/LINEAR (2002 LN13)*: Activity was first seen in this JFC by WISE 190 days after its perihelion while at a distance of 2.70 AU from the sun. No significant signal was observed in W1 or W2. A faint dust tail was apparent, as was a central condensation, easily separable from the dust, which yielded a nucleus size of ~2 km.

*233P/La Sagra (2009 WJ50)*: WISE viewed this Encke-type comet very close to its perihelion distance, at 1.81 AU from the sun, just 34 days before perihelion. The comet's faint tail indicated activity, and significant W2 excess that yielded a $CO_2$ production rate of $1.1 \times 10^{25}$ molecules per second. A strong central peak in the stacked image yielded an extracted nucleus flux corresponding to a size ~1 km.

*P/2009 WX51 (Catalina)*: WISE viewed this NEC 61 days after its perihelion (q = 0.8 AU), when it was at a distance of 1.26 AU from the sun. The comet displayed a tail, and significant W2 excess yielded a $CO_2$ production rate of $1.6 \times 10^{25}$ molecules per second. The W3 and W4 signals showed a strong central condensation, and the extracted nucleus flux yielded a size of <0.5 km.

*P/2010 B2 (WISE)*: The first WISE-discovered comet, 2010 B2 (WISE), was detected on 23 Jan 2010, just 32 days after its perihelion, at 1.64 AU, and again on 5 Aug 2010, at an outbound heliocentric distance of 2.49 AU. With a Jupiter Tisserand invariant ($T_J$) of > 3, and a semi-major axis less than Jupiter's, the comet was categorized as an Encke-type JFC. The first visit in January showed obvious activity, and detections in all four bands. The second visit showed significant signal only at 12 and 22 μm. The extracted flux from W3 and W4 in the first visit yielded nucleus size estimates, and showed W2 excess in the total signal. Our size estimate of 0.99 +/- 0.15 km implies the nucleus comprises less than a quarter of the total signal in the bands. The strength of the thermal signal in the second visit suggests the comet was still active at a distance of 2.5 AU, or that the dust component was still significant (~50% of the total signal), but did not have sufficient extended signal to remove the coma as was possible in the images from the first visit.

*P/2010 D1 (WISE)*: This comet, the second discovered by WISE on 17 February 2010, was detected only at one visit by WISE, 237 days after its perihelion, at a heliocentric distance of 3.02 AU. A faint coma and tail was shown in the stacked W3 and W4 images, making it identifiably a comet. However, extracted W1 and W2 signals were very faint, near the level of the noise, at or below 3-sigma. The extracted nucleus signal yielded a diameter of 2.5 km.

*P/2010 D2 (WISE)*: The third comet discovered by WISE was also a JFC, and it was detected only at one epoch by WISE curing the fully cryogenic mission, on 26 February, 2010. The images were taken very near to its perihelion, within 8 days, at 3.66 AU. As with P/2010 D1, the coma was faint, but present, yet lacked a distinctly extended tail. Removal of the coma signal yielded a nucleus size estimate of 4.65 km. In addition to strong flux in W3 and W4, the signal in the stacked images showed faint (5-sigma) signal in W1, but no significant signal in W2 that may have indicated $CO_2$ or CO emission down to production limits (1-sigma) of $3\times10^{25}$ and $3\times10^{26}$ molecules s$^{-1}$, respectively.

*C/2010 D3 (WISE)*: The first LPC discovered by WISE was detected at two separate epochs during the cryogenic survey. The first visit spanned 5 days centered around 26 February 2010 when the comet was 189 days prior to perihelion at a distance of 4.28 AU, and again outbound, 60 days after perihelion, when the comet was at a heliocentric distance of 4.52 AU. No significant W2 was detected in either visit, although weak W1 signal was seen during the first visit, allowing for $CO_2$ and CO production rate limits of $4\times10^{25}$ and $4\times10^{26}$ molecules s$^{-1}$, respectively. Faint coma is discernable in the stacked images of W3 and W4 during both visits, and the extracted nucleus signal yielded an estimated size of 4.3 km.

*C/2010 D4 (WISE)*: This LPC was discovered on 28 February 2010, 335 days after it perihelion, at a rather distant 7.43 AU from the sun. A second visit occurred over 5 July 2010, when the comet was at a larger heliocentric distance of 7.66 AU. The comet is not very active in the WISE images. No significant signal is present in W1 or W2, and the signal in W3 and W4 nearly match the WISE PSF, and are likely dominated by the nucleus. The 12 and 22 μm flux measurements from both visits yield a diameter of ~25km.

*C/2010 DG56 (WISE)*: When this LPC was discovered, at a distance of 1.95 AU from the Sun, 3 months prior to it perihelion, no signs of significant activity were apparent. The stack of its PSF-like images for this first visit yielded a diameter of 1.5 km. When the comet was imaged again at a heliocentric distance of 1.87 AU on 26 July 2010, the comet was quite active, enough to sufficiently obscure the nucleus, so that its signal could not be extracted from this second visit. However, though a slight W2 excess was possibly present, it was not above a 3-sigma uncertainty.

*C/2010 E3 (WISE)*:  Imaged when it was only marginally active, this LPC is on a parabolic orbit. The extracted nucleus signal yielded a size of 0.4 km, and no significant W2 or W1 signal was present. The comet was at a distance of 2.3 AU from the Sun at the time, very near its perihelion distance of 2.27 AU, which occurred one month after its discovery.

*C/2010 FB87 (WISE-Garradd)*: First observed inbound, the comet's image nearly matched the WISE spacecraft's PSF in W3 and W4, and showed no significant signal in W1 or W2.  These first observations were made when the comet was at a distance of 3.62 AU from the Sun, 224 days before perihelion. The comet was imaged a second time, still 108 days before its perihelion, while at a heliocentric distance of 3.04 AU.  A faint coma and tail were apparent in the images. WISE detected W1 and W2 signal during its second visit, but no significant W2 excess. However, when the comet was imaged by WISE a third time, in the post-cryogenic portion of the prime mission, W2 excess was present. The comet was then outbound at a heliocentric distance of 2.92 AU, 66 days following its perihelion at 2.84 AU.

*C/2010 G3 (WISE)*: This LPC was detected on two separate visits, both during the fully cryogenic phase of the WISE mission, and both while the comet was outbound. The first was within 4 days of perihelion, and the second 83 days after perihelion. With the furthest perihelion distance (4.9 AU) of the WISE/NEOWISE-discovered comets, one might expect CO or $CO_2$ to have driven the activity. Surprisingly, no significant W1 or W2 signal was seen at either visit, yet W3 and W4 images revealed both dust coma and tail. However, the dust modeling suggests that the activity occurred more than a year prior to the observations, and that potentially, the dust was composed of large-particles that lingered after the outburst. At such distances it was unlikely that $CO_2$, but rather that CO, was a possible driver. If so, the time of outburst, 726 days prior to perihelion, would have sufficiently preceded the WISE observations that the CO would have photo-dissociated. The predicted time scales for CO dissociation at 4.9 AU are ~370 days or less (Huebner et al. 1992), and ~90 days for $CO_2$.

*C/2010 J4 (WISE)*: Comet C/2010 J4 was detected on two visits in May of 2010, the first two days before its 3 May 2010 perihelion and the second 9 days after. This parabolic comet had a perihelion distance of 1.09 AU, and came within 0.31 AU of the Earth's orbit. Both sets of observations showed significant coma and dust tails. W2 signal was significant in the stacked images of both visits; however, no significant excess above the dust thermal contribution was seen. The dust signal heavily dominated the total signal, and the extracted nucleus flux and derived diameter should therefore be taken as an upper limit.

*P/2010 JC81 (WISE)*: This comet was detected twice during the WISE prime mission. The first visit was during the 4-band fully cryogenic period, when the comet was at a heliocentric distance of 3.9 AU, and the second was during the post-cryogenic period, when the comet was 2.65 AU from the sun. The first and second visits were 350 and 180 days, respectively, before the comet reached its perihelion at 1.8 AU from the sun, and before activity was noted by ground-based observations. The stacked images of both visits show a near-bare PSF-like surface brightness profile. The fitted temperature, about 40K above the black-body temperature, was consistent with a nucleus with a beaming parameter ~0.8. No strong indication of W2 excess was present in the fully cryogenic mission data, and the W1 and W2 fluxes in the second visit were also consistent with a bare nucleus with a beaming parameter near 0.8, and a visual-wavelength albedo on the order of a few percent. We note that it is possible for comets to be very active even out to large heliocentric distances, though small comets have been observed without coma as well. P/2010 JC81 should be an interesting candidate for study upon its return in 2034 for signs of a large nucleus.

*P/2010 K2 (WISE)*: This JFC was detected only once during the WISE fully cryogenic mission, and it exhibited a faint tail in W3 and W4. The images were taken when the comets was at a distance of 1.29 AU from the Sun, less than a tenth of an AU from its perihelion distance, and within 41 days of its perihelion passage. There was a clear W2 excess in the flux, which yielded a $CO_2$ production value of $1.3 \times 10^{25}$ molecules per second. The extracted nucleus flux was consistent with a sub-km size diameter.

*C/2010 KW7 (WISE)*: This LPC was imaged by WISE during the fully cryogenic mission 255 days prior to its perihelion and again 148 days before perihelion, at heliocentric distances of 3.7 and 3.0 AU, respectively. No significant W1 or W2 signal was seen in either data set. However, the W3 and W4 signals were strong, and the W3 and W4 brightness profiles matched WISE PSFs for the two band-passes. The data yielded a nucleus size for this body of ~6 km. Activity was later identified by observations taken at Spacewatch (c.f. Scotti, J.V., Williams, G. V. 2010. Comet C/2010 KW7 (WISE). Minor Planet Electronic Circulars 20.) 17 days following is perihelion at 2.57 AU from the Sun.

*245P/WISE (2010 L1)*: Activity was clearly shown in the first set of images WISE obtained of this JFC. The stacked images yielded no significant W1 or W2 flux values. The data were taken 120 days after the comet's perihelion, when it was at a distance of 2.6 AU from the sun. The nucleus signal extracted from the stacked images yields a diameter of ~1.5 km.

*C/2010 L4 (WISE)*: A dust coma and tail were apparent in the individual images of this LPC, taken 112 days before its perihelion while the comet was at a distance of 3 AU. The stacked images showed no significant signal in the two shortest band passes. The extracted nucleus signal corresponded to a diameter of 3.4 km.

C*/2010 L5 (WISE)*: The comet C/2010 L5 was the only WISE-discovered Halley type comet (HTC). It was strongly active when it was first detected 52 days after its perihelion, and again 85 days after. W2 excess was apparent in both of these visits, and yielded $CO_2$ production rates of $2.7 \times 10^{26}$ and $1.2 \times 10^{25}$ molecules per second, respectively. Because the comet was so active, the extracted nucleus signal, yielding a diameter of ~2 km, should be regarded as an upper limit. This is further supported by a non-detection at the comet's predicted location in January (Kramer et al. 2015). Large-grain dust modeling suggests that the comet's peak activity was near perihelion (Table 6). However, its worth noting that CO2 dissociation lifetimes are ~5 and 10 days for the comet's heliocentric distances of 1.2 and 1.6 AU, respectively, while $CO_2$ and CO lifetimes are ~22 and 39 days at these distances. It is unlikely, then, that outgassing had completely ceased very soon after perihelion. This particular case is discussed in detail in Kramer et al. 2015.

*P/2010 N1 (WISE)*: This inbound JFC was discovered at a heliocentric distance of 1.55 AU, 41 days prior to its perihelion. The comet was moderately active, but showed a strong W2 excess and yielded $CO_2$ production rates of $3.1 \times 10^{25}$ molecules per second. The extracted nucleus flux was consistent with a solid surface diameter of 0.9km.

*P/2010 P4 (WISE)*: The final comet WISE discovered during the prime mission was a JFC, detected 31 days after its perihelion. The stacked image revealed a faint tail with a morphology consistent with dust particles emitted long before perihelion. No significant W1 or W2 signal was detected, and the extracted nucleus signal matched a body with a diameter of 1.2km.

*C/2014 C3 (NEOWISE)*: As with all three comets discovered during the first year of the NEOWISE reactivated mission, no nucleus sizes could be confidently derived from the W1 and W2 images, owing to the level of activity and dust signal observed at these bandpasses. However, all three comets discovered by the reactivated NEOWISE mission to date showed 4.6 µm channel excess. This long-period comet showed W2 excess that yielded $CO_2$ production rates of $6.6 \times 10^{25}$ molecules per second at a heliocentric distance of 1.9 AU, 29 days after its perihelion.

*P/2014 L2 (NEOWISE)*: The second comet discovered during the reactivated NEOWISE mission was a JFC. NEOWISE imaged the comet 37 days before it reached perihelion, at a distance of 2.26AU from the sun. The stack images revealed a remarkably extended morphology in W2 relative to the more compact dust coma and tail in W1, and yielded a $CO2$ production rate of $2.4 \times 10^{27}$ molecules per second.

*C/2014 N3 (NEOWISE)*: This LFC was discovered by NEOWISE at a distance of 4.4 AU, 251 days prior to its perihelion, and it showed a faint tail and coma in the stacked images. The W2 excess yielded a corresponding $CO_2$ production rate of $2.7 \times 10^{26}$ molecules per second, although CO production at a rate of $2.9 \times 10^{27}$ molecules per second is the more likely driver at the distances the comet was observed. A second visit occurred 90 days before perihelion (Rh=3.96 AU), and apparent W2 excess yielded CO production rates of $3.9 \times 10^{27}$ molecules per second.

*P/2015 J3 (NEOWISE):* The latest NEOWISE comet discovery was made on May 15, 2015 at a distance of 1.67AU from the Sun. The JFC showed no indication of activity morphologically, but in ground-based follow-up images there was a faint tail. Size estimates for the nucleus are 2.3 +/- 0.7 kilometers, and reflectance 0.02 +/- 0.02, based on JPL's Horizons rough estimate of the visual-band nuclear magnitude at 18.5.

*2010 KG43:* This body, on a centaur-like orbit, was found to be active by Waszczak et al. (2013), but has not been confirmed as a comet. Significant flux values were observed in W3 (3.3 +/- 0.6 mJy) and W4 (9 +/-2 mJy) during the prime mission, yielding a preliminary diameter of 4 +/- 1 km, and an albedo of 0.02 +/- 0.02 for the object, assuming a beaming parameter near 1. This body is not included in the further analyses, since it's official cometary status remains undetermined.

*4.3 Nucleus Size*: The nucleus sizes of the discovery comets listed in Table 3 are shown as cumulative distribution plots in Figure 5. This sample is not large enough to make definitive conclusions as to the size frequency distribution power-law exponents for the total or sub-samples of LPCs and SPCs. Additionally, these samples are not de-biased in any way. However, because they were discovered by the WISE spacecraft, neither are they pre-selected by a strong visual reflectance bias. The data also includes two possible upper limits for the LPCs. However, these fall well below sizes that would influence the mean fractional sizes of the samples. An analysis using a minimum-variance unbiased estimator (MVUE) routine based on Feigelson and Babu (2012) was utilized to examine a power-law relation. The results were inconclusive, with the LPC sample yielding a size frequency distribution power-law exponent, $\alpha$, of 1.4 +/- 0.2, and the SPC sample yielding $\alpha$=1.6 +/- 0.2.

Figure 5 suggests that the LPC nuclei are, on average, larger than the SPC nuclei, by something like a factor of ~2. This conclusion is consistent with a similarly sized sample of LPCs presented in Lamy et al. (2004) compiled from sizes in the literature. Statistically, however, the sample sizes are small; a Kolmogorov–Smirnov test of the NEOWISE discovered LPC and SPC size distributions yields a 94% confidence that the diameters come from different distributions. One of the sources in the Lamy et al. (2004) compilation of diameters was Meech et al. (2004), which concluded from visual-wavelength data that there was no average size difference between LPCs and SPCs while using an assumed geometric albedo.  Note also that our sample contains SPCs with nuclei larger than 10 km in diameter. SPC nuclei on these scales were also measured in the SEPPCoN sample reported by Fernández et al. (2013).

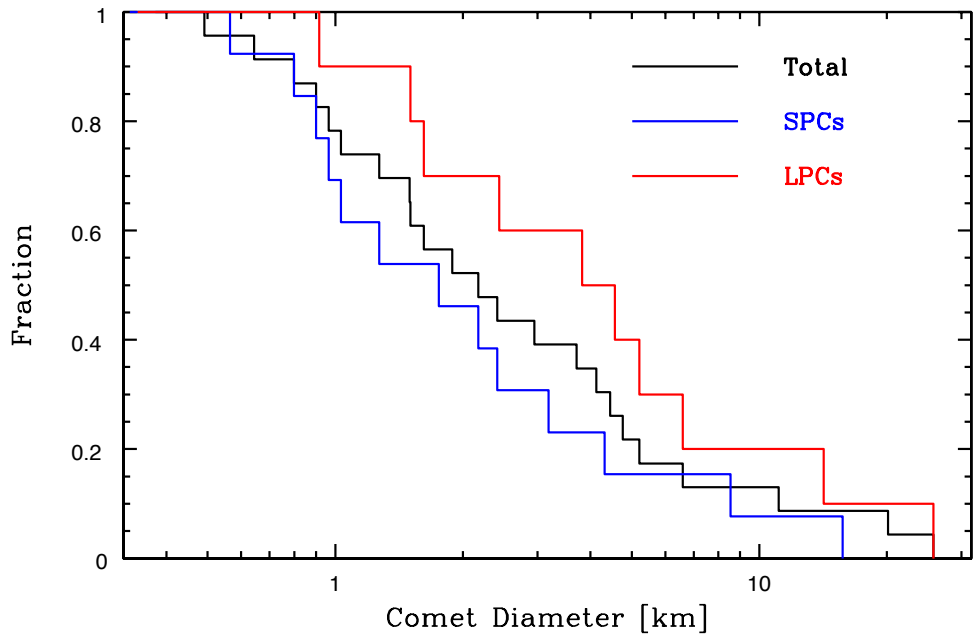

Figure 5: The cumulative size distribution of the nuclei of comets discovered by WISE/NEOWISE, including the 3 discoveries of activity (P/2002 LN31, P/2009 WX51, P/2009 WJ50), and P/2015 J3 from the NEOWISE reactivated mission, but not including the remaining 3 comets discovered during the reactivated mission since they had no 11 and 22 μm measurements and appeared active.

*4.4 Dust Tails:* Of the 21 comets discovered by WISE during its prime mission, 9 (6 LPCs and 3 SPCs) were found to have extended emission due to dust tails. We employed the well-described Finson-Probstein (Finson and Probstein, 1968) method to model these dust tails in order to constrain the size and age of the particles that comprised those tails. The Finson-Probstein method assumes that once a particle leaves the surface of a comet, its motion is only governed by solar radiation pressure and solar gravity, thereby allowing the particle motion to be parameterized using the ratio of these two forces, called $\beta$:

$$\beta = F_{rad}/F_{grav}$$

where $F_{rad}$ is the force due to solar radiation and $F_{grav}$ is the force due to solar gravity. Putting in the appropriate values for $F_{rad}$ and $F_{grav}$ and collecting the constant terms yields the ratio

$$\beta = \frac{CQ_{pr}}{\rho_d a}$$

where $\rho_d$ is the particle density [g cm$^{-3}$], $a$ is the particle radius in cm, $Q_{pr}$ is the scattering efficiency due to radiation pressure, and the factor $C = 5.78 \times 10^{-5}$ g cm$^{-2}$ comes from collecting all the constants into a single term. Thus, we can see that $\beta$ is inversely proportional to the size of the particle: larger $\beta$ values correspond to a smaller particle size, and vice versa.

The $\beta$ parameter is incorporated into the equation of motion, which is then integrated for a range of $\beta$ values using a numerical integrator (based on the work of Lisse et al., 1998, with the version used here described in further detail in Kramer 2014). For each comet, we ran the models for 5 years in 1-day increments, with $\beta$ values ranging from 0.0001 (roughly cm-sized particles) to 3.0 (sub-micron sized particles). This allowed us to fully explore the reasonable parameter space for each comet tail. The software returns a matrix of points that can be plotted as curves of constant emission date (synchrones) or curves of constant $\beta$ (syndynes). The models were over-plotted on each corresponding W4 image, allowing the best $\beta$ and time since emission to be found for each comet.

In order to determine the heliocentric distance at which strong emission occurred, we find the synchrone that most closely matches the brightest part of the tail, giving the number of days since the emission occurred. We then step back in the comet's orbit using the online tool *Horizons* from JPL to find the heliocentric distance of the comet at that time. The $\beta$ values listed in Table 6 do not mean that there were no small grains released, but more likely that either they have all been swept away already or that they are not optically active at W3 and W4 wavelengths. We further emphasize that this is not necessarily the only time that emission occurred for the comet; it is only where strong emission of particles which are still in the image frame occurred. Similarly for the interpretation of the syndynes, we note that the $\beta$ values listed in Table 6 correspond to the brightest part of the tail, and there are likely particles both larger and smaller than suggested by the best $\beta$.

The results of this method are shown in Table 6, as well as in Figure 6.

**Table 6: WISE/NEOWISE Discovery Comets Dust Model Summary**

| Name | Approx. $\beta$ | Approx. Em. (yrs) | Approx. Em. (days) | Days since Perihelion | Approx. $R_h$ at Em. (AU) |
|---|---|---|---|---|---|
| C/2010 DG56 (B) | 0.003 | 0.2 | 73 | 73 | 1.59 |
| C/2010 FB87 (B) | 0.01 | 0.25 | 90 | -107 | 3.46 |
| C/2010 G3 (A) | 0.001 | 2 | 730 | 4 | 7.3 |
| C/2010 G3 (B) | 0.001 | 2.25 | 821.25 | 84 | 7.35 |
| C/2010 J4 (A) | 0.003 | 0.1 | 30 | -2 | 1.21 |
| C/2010 J4 (B) | 0.01 | 0.1 | 30 | 9 | 1.14 |
| C/2010 L4 | 0.003 | 1 | 365 | 112 | 3.75 |
| C/2010 L5 (B) | 0.001 | 0.2 | 60 | 52 | 0.81 |
| C/2010 L5 (C) | 0.001 | 0.25 | 90 | 84 | 0.80 |
| 245P | 0.003 | 0.25 | 90 | 119 | 2.16 |
| P/2010 D1 | [1] | [1] | [1] | 238 | [1] |
| P/2010 P4 | 0.001 | 1 | 365 | 32 | 3.18 |
| P/2009 WX51 | [2] | [2] | [2] |  | [2] |
| 233P | 0.1 | 0.1 | 30 | -34 | 1.85 |
| 237P | [2] | [2] | [2] |  | [2] |

[1]: Tail is present, but too short or faint to make even make an estimate. [2] Orbit plane angle separation is too small to separate syndynes and synchrones.

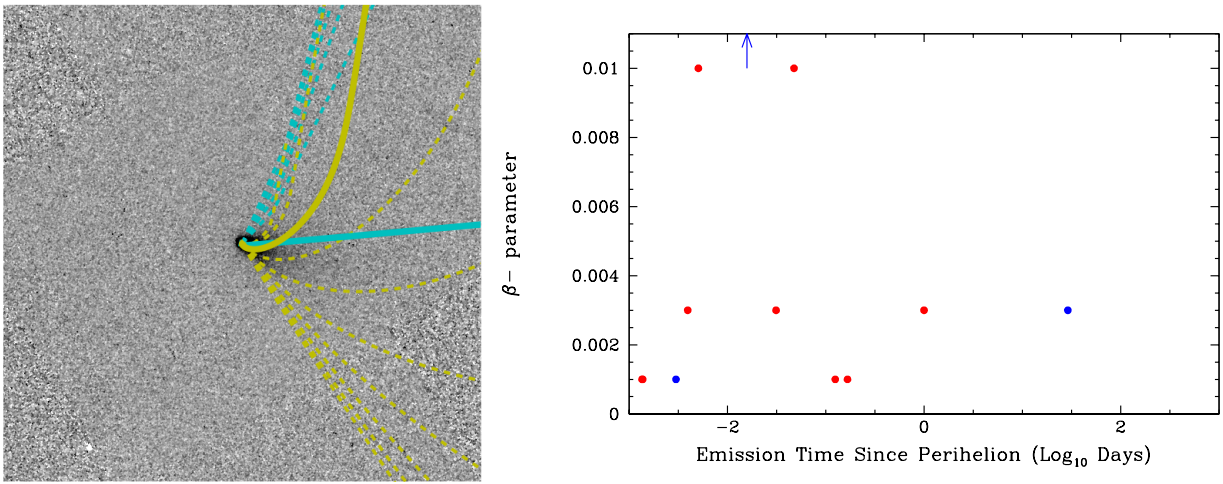

Figure 6: Panel A (LHS) – example of FP modeling for comet C/2010 L5 (WISE). Panel B (RHS) – Plot of days from perihelion verses best fit (by eye) of ejection time. Note no obvious trend is discernable. Note that the $\beta$ = 0.1 for 233P is indicated by the blue arrow near the top of the page.

*4.5 Dust Temperature:* Figure 7 shows the distribution of effective dust temperature based on the W3 and W4 band thermal fluxes. These may be effected by more than the dust temperature, namely variations in emissivity with silicate emission features (cf. Hanner et al. 2004). However, in a gross sense, these values are consistent with isothermal bodies with emissivity ~0.9. The standard deviation of the points from the thermal curve is +/- 17K.

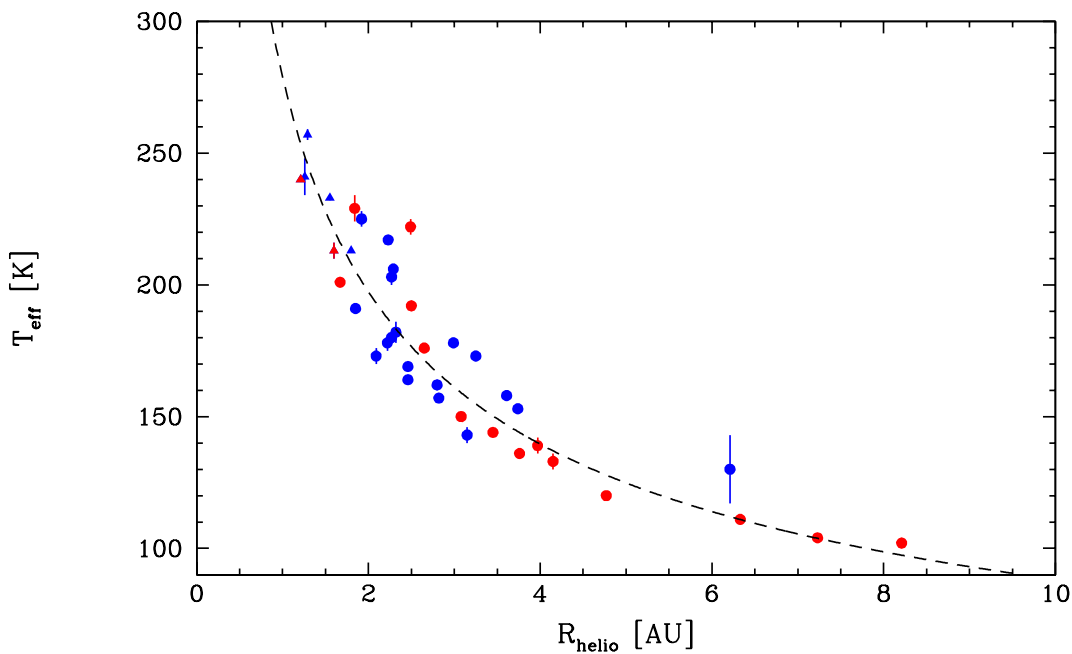

Figure 7: $T_{eff}$ is plotted as a function of Rh. As with previous plots, LPCs are indicated by read symbols, and SPCs by blue. The dashed line indicates the temperature of an isothermal body at the same distance from the sum with an emissivity of ~0.9.

*4.6 CO+CO$_2$:* The presence of CO or $CO_2$ manifests as a flux excess above the dust signal, as well as a difference in the morphology in W2 (Sections 2.3 and 3.3). WISE detected 163 comets during the prime WISE/NEOWISE mission. We found significant W2 flux excess in 40 comets, listed in Table 4. Assuming $CO_2$ was the dominant source of the W2 excess for all 40, we have converted the flux excess into CO2 production rate values (molecules per second), or $Q_{CO2}$. Of course, this is not valid for all the 40 comets. We know significant CO was detected for 29P (Senay & Jewitt 1994) and for C/2009 P1 (Garradd; Feage et al. 2014), observed at 6.2 and 6.3 AU, respectively, in the NEOWISE data. However, these values are readily convertible to approximate CO production rates ($Q_{CO}$) by multiplication of the ratio of $CO_2$ to CO fluorescence efficiencies ($g_{CO2}/g_{CO}$ = 11.6; c.f. Crovisier & Encrenaz 1983). The conversion to hypothetical $Q_{CO2}$ production rates facilitates possible comparisons between $CO_2$ and CO dominant behavior, and how it may be related to the quantity of dust present. We note, as discussed in Section 3.3, the listed uncertainties in the derived $CO_2$ production rates are the combination of the uncertainty in the calculated dust contribution as constrained by the W1, W3, and W4 photometry, added to the uncertainty from the W2 signal. Possible systematic sources of uncertainties, such as large variations in the fraction of CO relative to $CO_2$ or contributions to the W1 flux from non-dust signal, are not included in the tabulated values.

The $Q_{CO2}$ proxy and $\varepsilon f\rho$ values are plotted as a function of heliocentric distance in Figure 8. Except that the 15 LPCs may be more active than 24 SPCs at heliocentric distances greater than 4AU, no differentiating trend is readily apparent for $CO_2$ production. To identify correlations, a Kendall-$\tau$ test was applied to the LPC and SPC distributions for comet heliocentric distance with $CO_2$ production and with $\varepsilon f\rho$; high $\tau$ values near 1 indicate a correlation between the two parameters, and low values (~-1) indicate anti-correlations, where values near zero indicate no correlation. A second variable, the two-sided probability parameter, or *p-value*, is a test for a null hypothesis, such that a low *p-value* indicates a higher likelihood of the result indicated by the $\tau$ value. For example, a well-correlated pair of parameters, but poorly sampled, would have $\tau$ and *p-value* close to 1, and well-sampled correlated pair would have $\tau$ ~1 and the *p-value* ~0, but a well-sampled, but random, pair of parameters would have both values close to zero. Inside of 4 AU, the Kendall-$\tau$ test yielded $\tau$ values near 0 (0.18 and -0.10 for LPCs and SCPs, respectively), with low significance (*p*-values of 0.39 and 0.57 for LPCs and SPCs, respectively). Both SPCs and LPCs appear to have similar distributions given the limited sample. This find, in and of itself is a significant constraint on current solar system formation theories. Dones et al. (2004), for example, place the source of Oort cloud comets and KBOs to be near Jupiter and near Neptune, respectively. However, A'Hearn et al. (2012) suggests that volatile abundances are similar for differing dynamical classes of comets, implying a comparable formation environment between the CO and $CO_2$ sublimation zones. This data set may affirm this notion, which places profound constraints on the various solar system scenarios, specifically regarding planetary migration. Such theories (cf. Walsh et al. 2011, Morbidelli et al. 2008) which have previously suggested different formation regions for cometary types, may have to account for the comparable compositional profiles between differing comet dynamical populations.

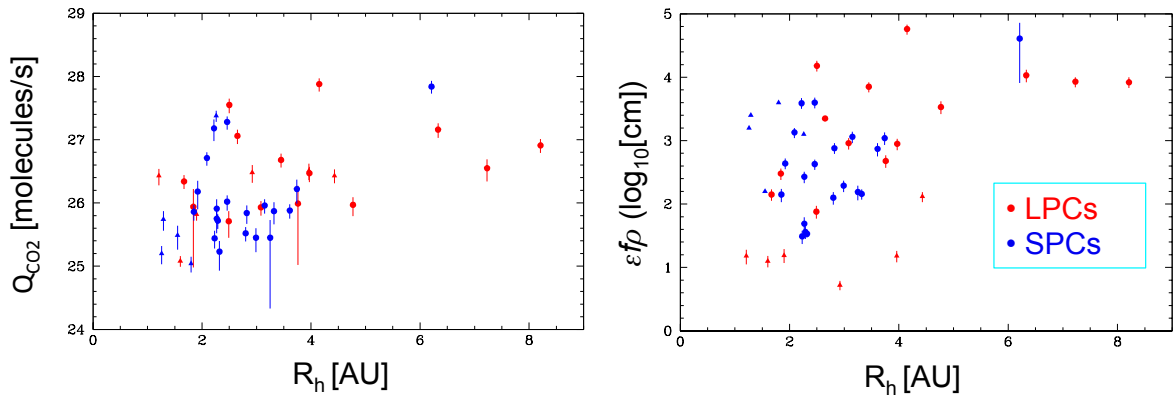

Figure 8: Panel A (LHS) excess flux in 4.6 μm channel converted to CO2 production plotted WRT heliocentric distance. Panel B (RHS) $\varepsilon f\rho$ as a function of heliocentric distance. Note the distributions for LPCs and SPC are similarly scattered.

Unsurprisingly, $\varepsilon f\rho$ values correlate well with SPCs ($\tau$=0.5, *p-value* = 0.02) and LPCs ($\tau$=0.4, *p-value* = 0.02) alike, as suggested previously (cf. Kelley et al. 2013). When the two values are ratio-ed ($\log_{10} Q_{CO2} - \log_{10} \varepsilon f\rho$), as in Figure 9, we find a relation of $R_h^{-2}$. Inverse proportionality with $R_h$ have been seen with OH and CN (A'Hearn et al. 1995), yet the relationship with $CO_2$ is particularly clear. Furthermore, where the A'Hearn et al. (1995) gas-to-dust ratio appears to go as $R_h^{-1/2}$, what we find with respect to $CO_2$ gas is considerably steeper. Such behavior, whereby an increase in $CO_2$ gas production lacks a corresponding increase in dust production, was noted in 103P (A'Hearn et al. 2011). This relation persists out to 4AU, where the trend deviates. If $CO_2$ is expected to drive activity more within these ranges of $R_h$, this may indicate that the bulk of $CO_2$ may be endogenic with the dust. Alternatively, or possibly concurrently, it may be that CO reaches its maximum production before 4AU if it resides at depth and its sublimation is not driven directly by surface insolation of, say, near-surface CO ice, while $CO_2$ is.

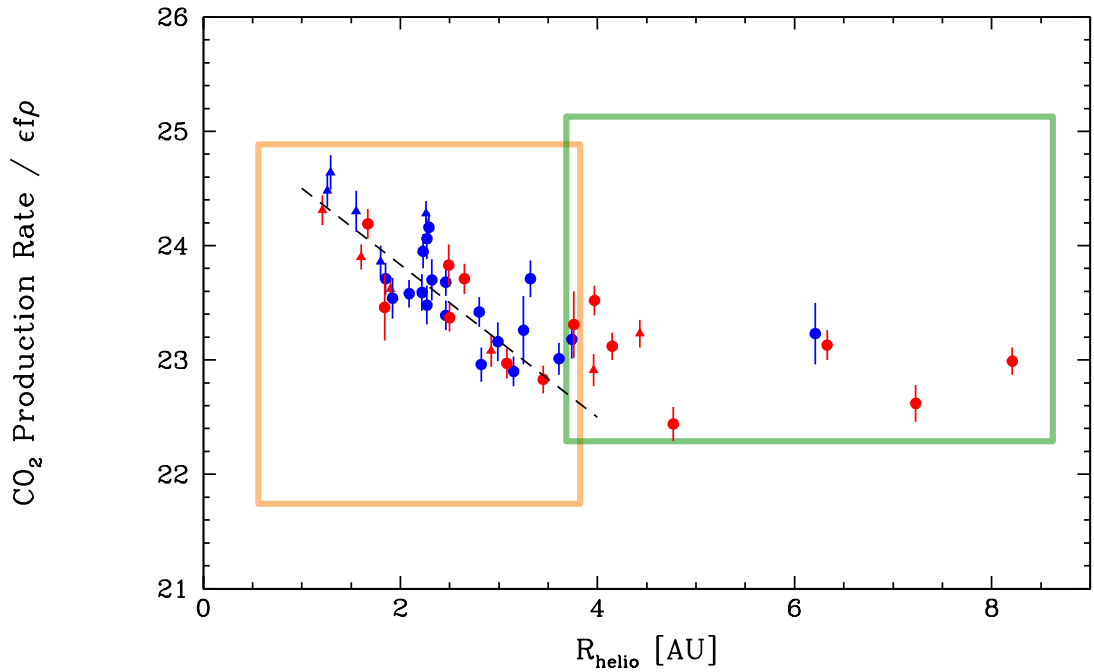

Figure 9: Log ($Q_{CO2}/\varepsilon f\rho$) as a function of $R_h$[AU]. LPCs are represented by red symbols, SPCs blue. Note the two dissimilar groupings of behavior (orange an green boxes) inside and outside 4 AU. The latter may be CO production driven activity, and includes 29P at 6.2 AU, and C/2009 P1 (Garradd) at 6.3 AU.

We found limited overlap between the Akari (Ootsubo et al. 2012; hereafter O12) and Spitzer Space Telescope (Reach et al. 2013; hereafter R13) observations. A total of 13 objects shared reported observations that allowed comparisons with the NEOWISE sample's somewhat larger time intervals. These results are summarized in Figure 10. The fidelity of the comparisons, however, are somewhat limited in several respects. Neither the R13 nor O12 observations provided $\varepsilon f\rho$ values, so that only gas production rates could be compared. Furthermore, O12 provided spectrally derived relative abundances, information we did not have, so that in order to make comparisons with our values, we converted the CO and $CO_2$ rates into proxy $CO_2$ production rates by dividing the O12 CO rate by 11.6, the scaling factor between the line strengths, and adding it to the reported $CO_2$ rate. Also, many of the comets that overlapped nonetheless were observed at similar times and heliocentric distances. Finally, we did not compare the limits of the non-detections in O12 and R13, but only detections. We found that, similar to the behavior for other species as analyzed in A'Hearn et al. (1995), variations in individual comet behavior did not clearly indicate trends with heliocentric distance for our proxy $CO_2$ values. What we found with the aggregate total sample shown in Figure 8, with dispersed behavior, seemed to match with the stochastic nature of cometary emission seen in Figure 10.

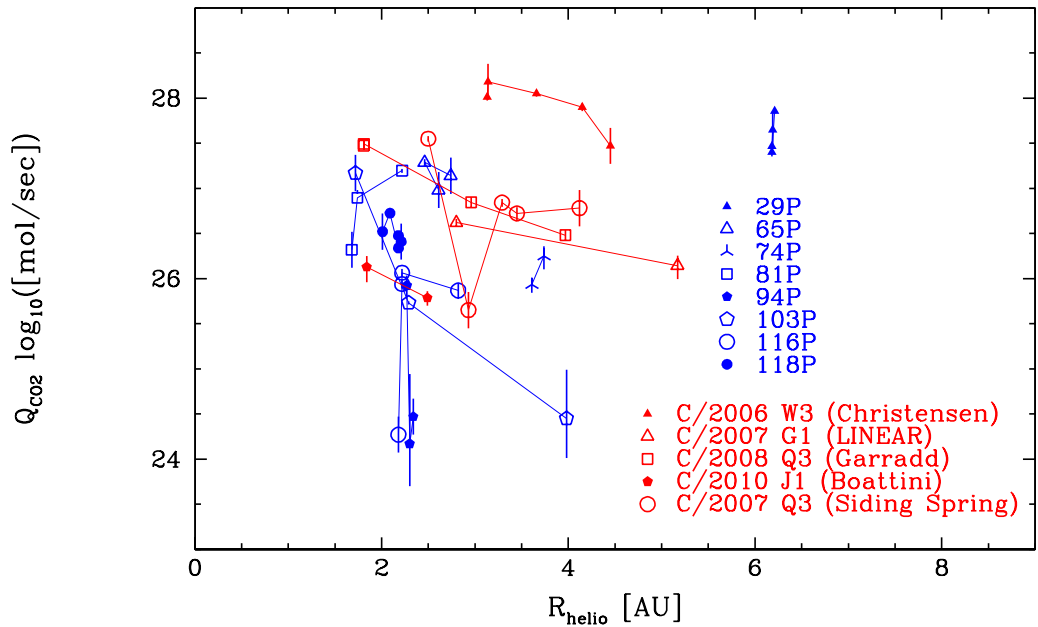

Figure 10: The comparative production rates of $CO_2$ as a function of heliocentric distance in individual comets. SPCs (blue) and LPCs (red) observed by Akari (Ootsubo et al. 2012), Spitzer Space Telescope (Reach et al. 2013) and WISE/NEOWISE are shown with distinguishing symbols, and with their data points connected. The Akari CO and $CO_2$ production rates were converted to proxy $CO_2$ rates for comparison with the Spitzer and WISE data sets (See text).

# 5 Conclusions

The 25 NEOWISE cometary discoveries are a relatively small but important and representative sample of comets detected by WISE/NEOWISE. We find the following from our analysis of this smaller sample:

- Long period comet nuclei may be, on average, larger than small period comet nuclei. Though the evidence from the sample is suggested, it must be confirmed by later efforts to account for the effects of biases as a function of orbital elements and size, as well as using the larger expanded sample from the WISE data.
- Dust detected at longer thermal wavelengths is large, often up to sizes of millimeters. Few comets reach peak activity after perihelion.

A total of 39 comets out of 163 detected by WISE/NEOWISE showed W2 excess, comprising nearly a quarter of the total sample detected in the WISE prime mission data. Our analysis of the sample of active comets, which have dust temperature constraints and differ morphologically in the 4.6 $\mu$m band, suggests:

- There is little difference between the nature of the dust production of LPCs and SPCs as a function of heliocentric distance.
- Similarly, the distribution of CO or $CO_2$ production as a function of heliocentric distance looks comparable for LPCs and SPCs, though fractionally more LPCs may be producing CO or $CO_2$ at heliocentric distances greater than 4 AU. The appearance of more LPCs exhibiting CO+$CO_2$ at these greater distances may suggest an evolutionary effect, such that LPCs retain

   their more volatile CO, while both LPCs and SPCs may have on average similar $CO_2$ abundances.

- Temperatures of dust seen at 12 and 22 μm are to first order well-approximated by an isothermal black body with emissivity ~0.9 and with a temperature range within +/- 17K (1-$\sigma$ dispersion).

- The ratio of CO or $CO_2$ production to the quantity of dust observed ($\varepsilon f\rho$) may follow a relation of ~$R_h^{-2}$ within 4AU. No similar relation seems to persist for greater distances. This may be attributable to different source regions (surface vs. sub-surface) for cometary CO and $CO_2$ emissions.

## Acknowledgements

This publication makes use of data products from the Wide-field Infrared Survey Explore, which is a joint project of the University of California, Los Angeles, and the Jet Propulsion Laboratory/California Institute of Technology, funded by the National Aeronautics and Space Administration. This publication also makes use of data products from NEOWISE, which is a project of JPL/Caltech, funded by the Planetary Science Division of NASA. This material is based in part upon work supported by the NASA through the NASA Astrobiology Institute under Cooperative Agreement No. NNA09DA77A issued through the Office if Space Science. R. Stevenson and E. Kramer were supported by the NASA Postdoctoral Program, and E. Kramer acknowledges her support through the NASA Earth and Space Science Fellowship program. We thank the Astrophysical Journal Editor for the very helpful comments regarding manuscript drafts, and the anonymous reviewer for providing valuable comments, both of whom greatly improved the paper content. The lead author also benefited greatly from a discussion with Nader Haghighipour of the Institute for Astronomy and NASA Astrobiology Institute, University of Hawaii-Manoa.